\documentclass[10pt]{elsarticle}
\usepackage{lineno,hyperref}
\usepackage{geometry}
\usepackage{setspace}
\usepackage{amsthm}
\usepackage{amsmath,bm}
\usepackage{amsfonts}
\usepackage{amssymb}
\usepackage{appendix}
\usepackage{bm}
\usepackage{color,colortbl,soul}
\usepackage{multirow}
\usepackage{graphics,graphicx}
\usepackage{caption}
\usepackage{epstopdf} 
\usepackage{eurosym}
\usepackage{float}
\usepackage{fancyhdr}
\usepackage{longtable}
\usepackage{mathrsfs}
\usepackage{enumerate}
\usepackage{flushend}
\usepackage[framemethod=tikz]{mdframed}
\usepackage{lipsum}
\usepackage{float}
\usepackage{verbatim}
\usepackage{cases}
\usepackage{upgreek}
\usepackage{subcaption}
\usepackage{hyperref}
\usepackage{booktabs}
\usepackage{commath}
\usepackage{multicol}
\usepackage{mathabx}
\usepackage{xcolor}
\usepackage{framed}
\usepackage{nomencl}
\usepackage{natbib}
\usepackage{tabu}
\usepackage[noend]{algpseudocode}
\usepackage[ruled]{algorithm2e}
\makenomenclature
\SetKwInOut{Input}{input}
\SetKwInOut{Output}{output}
\journal{Computers \& Structures}
\date{}
\makeatother
\setcitestyle{authoryear}
\usepackage{times}
\begin{document}
\begin{frontmatter}
\title{
Characterizing failure morphologies in fiber-reinforced composites via $k$-means clustering based multiscale framework}
\author[add1]{Harpreet Singh\corref{corrauth}}
\cortext[corrauth]{Corresponding author}
\ead{harpreet@iitgoa.ac.in}
\address[add1]{School of Mechanical Sciences, Indian Institute of Technology Goa, Farmagudi, Goa 403401 India}		
\begin{abstract} \begin{spacing}{1.1}\small
A novel homogenization methodology is proposed for analyzing the failure of fiber-reinforced composite materials, utilizing elastic and eigen influence tensors within a damage informed transformation field analysis (D-TFA) framework. This approach includes a technique for calculating macroscopic damage under uniform stress and strain conditions, offering more realistic simulations. Computational efficiency is enhanced through a reduced-order modeling strategy, while elastic and eigen strain distribution driven $k$-means clustering methods are employed to partition the microscale domain. The model's performance is assessed by simulating the response of a representative volume element (RVE) treated as a homogenized continuum. Subsequently, a comparative \color{black}assessment \color{black}  is carried out to check the efficacy of two clustering schemes. Damage morphologies are calculated using proposed framework and compared with predictions obtained using finite element method. Furthermore, open-hole specimen tests are simulated and failure paths are predicted for the domains with different fiber layups. Ultimately, we show that D-TFA can accurately capture damage patterns and directional strengths, providing improved predictions of the mechanical behavior of composite materials. It has been demonstrated that higher cluster counts are crucial for capturing a more accurate stress-strain response, especially for complex microstructures.\end{spacing} 
\end{abstract}

\end{frontmatter}

\section{Introduction}

In the last two decades, composite materials are aggressively being used in aerospace, sports, defence, automotive, and medical industries, offering high stiffness and strength with low mass. The mechanical behaviour of the composite structure is complex due to its inherent heterogeneous microstructure and anisotropic response. Furthermore, the failure of composite material is difficult to predict, which limits its application in structures. Although numerous efforts were carried out to study the progressive failure of composites using macroscopic constitutive models, anticipating the onset of failure and its evolution is still a challenge to overcome. These single-scale macroscopic frameworks depend on the loading, which often fails to predict failure mechanisms for various multiaxial loading conditions.
On the other hand, the multiscale models transverse from microscale (scale of fiber-matrix) to the macroscale (scale of laminate) and demonstrate their capabilities to predict mechanical behaviour when subjected to complex loading scenarios. Recognizing that the failure starts at the microscale, many researchers have utilized the multiscale approaches to obtain the failure characterisation of composite laminates. By capturing the microscopic failure modes, i.e. matrix cracking, fibre breakage and interfacial debonding, the macroscopic failure architecture in terms of crack trajectories and overall stiffness reduction can be envisaged. 

In order to simulate the laminate mechanical response using multiscale models, researchers have proposed several frameworks (see  \citet{Feyel2000},  \citet{Feyel2003}, \citet{Raghavan2004}, \citet{E2007}, \citet{Belytschko2008} and more). Largely, the models devised during the last three decades utilize 1). the finite element method (FEM) to characterize the microstructure and 2). a computational homogenization-based framework to couple the micro response with macroscale behaviour. \color{black}These computational homogenization-based multiscale methods have the capability to include the effects of microstructure failure either by appending fracture-mechanics based techniques (see  \color{black}\citep{verhoosel2010computational}, \citep{vsmilauer2011multiscale},  \citep{coenen2012multia}, \citep{coenen2012multi}, \citep{greco2012non},  \citep{bosco2014multiscale}, \citep{toro2016cohesive}) or by employing the continuum damage mechanics based formulations  (see  \citep{fish1999computational}, \citep{fish2001multiscale}, \citep{oliver2004continuum}, \citep{oliver2008two}, \citep{oliver2014crack}, \citep{bogdanor2015multiscale}). \color{black}The fracture-mechanics methodologies utilize a discontinuous displacement field representing the cracked surface in the material. The information marches from micro to macroscale by maintaining the equivalence of dissipated energies across the scales. In contrast, the continuum damage mechanics-based approaches exploit each micro-phase as a statistically homogeneous medium, endowing an effective material law and an internal variable that institutes the state of damage accounting for the stiffness degradation. Among these two techniques, the damage mechanics approaches are more alluring due to their low computational cost and reasonably less solution time. 

To evade the issue of computational complexity, the transformation field analysis (TFA) based methodologies have been adopted vastly by several researchers. Fundamentally, TFA employs a reduction of order approach where the full field information is condensed into a small number of sets. Each set uniquely represents a subregion of the microdomain and establishes the microdomain characterisation in terms of concentration and influence functions. Despite the low computational time, the original version of TFA failed to account for the damage-induced effects and could not accurately predict the fracture trajectories. Furthermore,  several improvements in TFA were executed to capture the macroscale in-elastic deformations caused by the plasticity of micro-constituents (\citet{sacco2009nonlinear}, \citet{fritzen2011nonuniform}, \citet{sepe2013nonuniform}, \citet{marfia2018multiscale}, \citet{gopinath2018common}). In order to capture the stiffness degradation due to progressive damage,  \cite{fish1997computational} introduced a TFA-based approach where the effect of eigenstrains, representing inelastic strains, thermal strains, and/or phase transformation strains are accounted. Later on, a closed-form solution for determining macroscale fields from local solutions was developed by \cite{fish1999computational} leveraging asymptotic analysis and homogenization theory to model damage in periodic microstructures.  \cite{oskay2007eigendeformation} extended this approach by incorporating interphase failures through the concept of eigendeformation. However, a major limitation of the lower-order TFA approximation is the emergence of spurious post-damage stiffness. To address this issue, techniques such as reconstructing the eigenstrain field using recomputed influence tensors \citep{fish2013hybrid} or employing enhanced stress-strain laws for subdomains \citep{singh2020strain} have been proposed. The eigendeformation-based TFA has been widely applied to analyze the response of fractured multi-phase media, as demonstrated in the works of \citep{crouch2010symmetric},  \citep{bogdanor2013uncertainty}, \citep{bogdanor2015multiscale}. \cite{zhang2017sparse} developed an enhanced eigenstrain-based method to calculate macroscopic responses of polycrystalline materials and improved computational efficiency by introducing sparsity through an advanced grain clustering scheme. In order to reduce the computational cost, \cite{oskay2020discrete} presented an eigendeformation-based reduced-order homogenization technique for failure analysis of composite materials, incorporating cohesive surfaces to model microscale failure evolution.  Recently, \cite{su2022modeling} developed a multiscale discrete damage theory (MDDT) to simulate randomly oriented and progressively reoriented cracks in cross-ply laminates, achieving mesh size-objectivity through a length scale parameter. This model captures multiple failure modes like splitting, transverse matrix cracking, and delamination. In an another research work, \cite{addessi2023non} utilized a reduced-order non-uniform TFA-based multiscale model to analyze masonry material modeled with shell elements, considering out-of-plane and in-plane loading. Interface failure was characterized using cohesive-frictional zero-thickness contact surfaces. Despite numerous studies utilizing various TFA versions for capturing the failure of composites, several key challenges still persist:

\begin{enumerate}
	\item \textbf{Optimal approximation of damage-equivalent eigenstrain fields:} Determining the optimal number of reduced order variables remains a critical issue, as explored by \citep{sparks2013identification} and \citep{bogdanor2015multiscale}.
	\item \textbf{Spurious eigenstrain field and macroscale response:} The unrealistic post-damage stiffness associated with lower-order TFA approximations can lead to inaccurate predictions. Researchers like \citep{fish2013hybrid},  \citep{singh2017reduced}, and \citep{singh2020strain} investigated methods to address this problem.
\end{enumerate}
These observed issues are primarily attributed to the partitioning map of the microdomain, which is generated by the order reduction technique. 

 Recent research has introduced new order-reduction methods for minimizing the computational complexity in analyzing materials with complex microstructures. These machine learning-inspired data-driven clustering-based homogenization techniques primarily include self-consistent clustering analysis (SCA) (\cite{Liu2016, Liu2018a, Kafka2018, Liu2018b, Li2019, Shakoor2019, Yu2019}), virtual clustering analysis (VCA) (\cite{tang2018virtual, zhang2019fast, zhu2021adaptive, yang2022comparative, yang2023virtual}), FEM-based clustering analysis (FCA) (\cite{cheng2019fem, nie2019principle, nie2021efficient}) and cluster-based nonuniform transformation field analysis (CNTFA) (\cite{ri2021cluster}). \cite{Liu2016, Liu2018a, Liu2018b} proposed SCA for dividing the material into clusters based on the strain concentration tensor and used \textit{k}-means clustering for grouping the data. The macroscopic stress and strain of the material are then expressed in terms of the interactions between these clusters. A novel aspect of SCA is its data-driven clustering approach which groups elements within the material based on their strain levels under elastic loading, assuming similar behaviour under subsequent loading. SCA employs the Lippmann-Schwinger equation to determine the effective nonlinear properties of the material. This equation relies on Green's function for an elastic reference material, which is updated iteratively using a self-consistent scheme. The interaction matrices, representing the influence of one cluster on another, are efficiently computed using the fast fourier transform (FFT). Green's function for isotropic linear elastic materials is particularly convenient for solving the incremental Lippmann-Schwinger equation. On the contrary, VCA identifies and groups similar microstructural features into simplified virtual clusters, treating them as representative of the actual microstructure (\cite{tang2018virtual, zhang2019fast, zhu2021adaptive, yang2022comparative, yang2023virtual}). Boundary value problems are then solved on these virtual clusters to calculate their effective properties. For complex microstructures, VCA often provides more accurate results. However, SCA can be computationally more efficient for simpler materials. It is formulated in a consistent framework of the finite element method. FCA is another clustering-based computational method for analyzing heterogeneous materials (\cite{cheng2019fem, nie2019principle, nie2021efficient}). It divides the microstructure into clusters, constructs an interaction matrix using FEA, and applies eigenstrain to microdomains. Unlike SCA, FCA doesn't require a reference material or Lippmann-Schwinger equation. It uses a principle of minimum complementary energy and statically admissible cluster stress fields for nonlinear analysis. Alternatively, CNTFA simulates the behaviour of heterogeneous materials based on local constitutive relations, eliminating the need for preliminary assumptions or approximate expressions for effective behaviour (\cite{ri2021cluster}). Clustering of the domain is performed by determining the inelastic response of microstructure subjected to a uniform strain field over the boundaries.    
 
Inspiring from this approach, the present manuscript proposes a novel clustering driven damage informed TFA methodology for obtaining the mechanical response of fiber-reinforced composites in the presence of damage. To achieve order reduction in the multiscale homogenization procedure, the RVE domain is divided into subdomains with constant eigenstrain fields analogous to the piecewise uniform TFA method. The clustering map for microdomain is obtained using eigen strain snapshots. Furthermore, this approach incorporates a novel methodology for the calculation of macroscopic damage for uniform stress and strain conditions, which leads to more realistic computations by addressing the issue of post-damage stiffness. By leveraging elastic and eigen influence tensors, the calculations of homogenized properties are performed, significantly reducing the computational effort required for preprocessing. Microscale studies are performed to predict the damage morphologies and directional stiffness degradations. Eventually, the numerical results are compared with experimental data to assess the performance of the proposed TFA-based homogenization technique.

The subsequent sections of this paper are organized as follows: In Section \ref{sec:2}, multiscale philosophy in the context of the homogenization framework is introduced, and coupling between scales is explained. Additionally, section \ref{sec:2} explains partitioning based reduction of order technique. the order reduction methods focuses primarily on the microscale model and elastoplastic damage constitutive laws followed for phase materials. Section \ref{sec:3} illustrates the D-TFA-based multiscale model and elaborates on the methodology for calculating the different influence and damage tensors for accounting failure in the material. $k$-means \color{black}clustering \color{black} scheme for domain partitioning is presented in section \ref{sec:4}. Section \ref{sec:5} details the numerical procedure used to implement the proposed method, covering both preprocessing and solution stages. Numerical studies, including RVE simulations and full-scale analyses of open-hole laminates are presented in Section \ref{sec:6} to validate the proposed approach. Section \ref{sec:7} summarizes the conclusions.                    
\subsection*{Notation}

\begin{itemize}
	\item \textbf{Scalars}: Regular Greek or Latin lowercase letters (e.g., $\beta$, $a$)
\item \color{black}\textbf{Vectors:} Boldface Latin lowercase letters (e.g., $\mathbf{\color{black}u}$, $\mathbf{v}$)
\item \textbf{Second Order Tensors:} Boldface Greek letters (e.g., $\boldsymbol{\sigma}$, $\boldsymbol{\epsilon}$, $\mathbf{\color{black}\Lambda}$, $\boldsymbol{\Phi}$)
 \item \textbf{Fourth Order Tensors:} Boldface Latin uppercase letters (e.g., $\mathbf{A}$, $\mathbf{B}$)
\item \textbf{Dot product of vectors:} $\mathbf{u} \cdot \mathbf{v} = u_i v_i$ (Einstein summation convention)
\item \textbf{Double contraction of second order tensors:} $\mathbf{\color{black}\Lambda} : \mathbf{\Phi} = \Lambda_{ij}\Phi_{ij}$ or $\boldsymbol{\epsilon}:\boldsymbol{\sigma} = \epsilon_{ij}\sigma_{ij}$
\item \textbf{Double contraction of fourth order tensors:} $\mathbf{A} : \mathbf{B} = A_{ijmn}B_{mnkl}$ 
\item \textbf{Double contraction of fourth and second order tensors:} $\mathbf{A} : \boldsymbol{\epsilon} = A_{ijkl}\epsilon_{kl}$ 
\item \textbf{$\mathcal{L}_2$ norm:} $\|\mathbf{u}\|_2 = \sqrt{\sum_{i=1}^n u_i^2}$
\item \textbf{Superscript in parentheses:} Specific partition number (e.g., $\boldsymbol{\sigma}^{(i)}$)
\item \textbf{Subscript in parentheses:} Time or load step number (e.g., $\boldsymbol{\sigma}_{(n+1)}$)
\item \textbf{Subscript in square brackets:} Iteration number (e.g., $\boldsymbol{\sigma}_{[m+1]}$)
\item \textbf{Superscript in square brackets at the start:} Load case number (e.g., $^{[{i}]}\bar{\boldsymbol{\sigma}}$)
\end{itemize}
\section{Reduced Order Homogenization}\label{sec:2}
The homogenization framework provides coupling between microscopic and macroscopic fields. The macro-strain is known at each point in the macroscale continuum domain, while the macro-stress field is computed by solving a boundary value problem at the microscale.

\begin{figure}[H]
\centering
\includegraphics[width=1\textwidth]{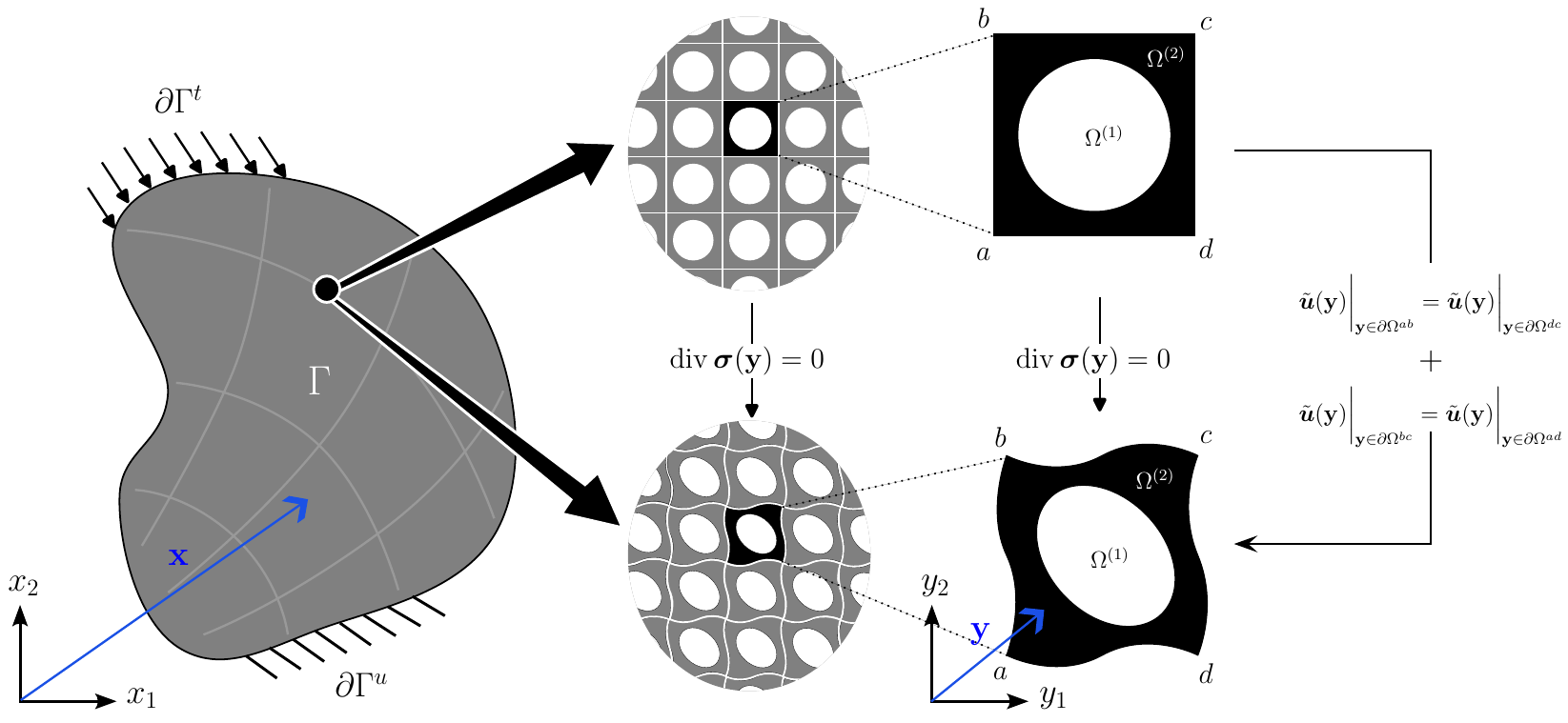}
\caption{Macroscale boundary value problem where \(\boldsymbol{x}\) denotes a position in the macroscale domain \(\Gamma\). \(\partial\Gamma^u\) and \(\partial\Gamma^t\) represent the Dirichlet and Neumann boundaries, respectively. For any point in the macro-domain, there exists a representative volume element with domain \(\Omega\) (marked \(abcd\)) with position representation as \(\mathbf{y}\). \(\Omega^{(1)}\) and \(\Omega^{(2)}\) illustrate the heterogeneous nature of the micro-domain.}
\label{fig:1}
\end{figure}

Each macroscale material point within the domain \(\Gamma\) is associated with a Representative Volume Element (RVE) defined by the domain \(\Omega\) and its boundary \(\partial\Omega\). This RVE encapsulates the microstructure of the material (refer to Fig. \ref{fig:1}). The averaging theorems establish the connection between macroscopic and microscopic quantities and allow macroscopic field variables to be expressed in terms of volume-averaged microscale field variables. When considering the stress field, this relationship can be formulated as
\begin{equation}
	\boldsymbol{\sigma}^o=\langle \boldsymbol{\sigma} \rangle=\dfrac{1}{\vert{\Omega}\vert}\int_\Omega \boldsymbol{\sigma}(\mathbf{y})\, d\mathbf{y}  \label{eq:1}
\end{equation}
where \(\boldsymbol{\sigma}^o\) represents the macrostructural stress tensor at a position \(\mathbf{x}\) within \(\Gamma\), and \(\boldsymbol{\sigma}(\mathbf{y})\) denotes the microstructural stress field within RVE under surface tractions \(\boldsymbol{\sigma}^o \boldsymbol{n}\Big\vert_{\partial \Omega}\) with position representation as $\mathbf{y}$. Here, \(\boldsymbol{n}\Big\vert_{\partial \Omega}\) denotes the outward normal to the boundary.

Similarly, the average strain field is expressed as
\begin{equation}
	\boldsymbol{\epsilon}^o=\langle \boldsymbol{{\epsilon}} \rangle=\dfrac{1}{\vert{\Omega}\vert}\int_\Omega \boldsymbol{{\epsilon}}(\mathbf{y})\, d\mathbf{y} \label{eq:2}
\end{equation}
where \(\boldsymbol{\epsilon}^o\) represents the macrostructural strain tensor at a position \(\boldsymbol{x}\) in \(\Gamma\), and \(\boldsymbol{\epsilon}(\mathbf{y})\) denotes the microstructural strain field within the RVE when boundary displacements are applied as \(\boldsymbol{\epsilon}^o \mathbf{y}\Big\vert_{\partial \Omega}\). The position vector for the boundary is indicated by \(\mathbf{y}\Big\vert_{\partial \Omega}\).

The RVE may contain either a single heterogeneity or multiple heterogeneities of the same or different materials embedded within another material. For fiber-reinforced composites, an RVE domain $\Omega$ constitutes a fiber phase as subdomain \(\Omega^{(1)}\) and the matrix phase as subdomain \(\Omega^{(2)}\) (see Fig. {\ref{fig:1}}) without any material discontinuity between \(\Omega^{(1)}\) and \(\Omega^{(2)}\). This serves as a unit cell for a periodic arrangement of heterogeneities that generates the full domain by repeating this unit cell. The displacement field for this microscale domain is denoted by \(\boldsymbol{u}(\mathbf{y})\) with components \(\{u_1, u_2, u_3\}\). The stress field \(\boldsymbol{\sigma}(\mathbf{y})\) can be calculated by solving the equilibrium equation for the RVE;
\begin{equation}
	\text{div}\hspace{1mm}{ \boldsymbol{\sigma}(\mathbf{y})} = 0 \label{eq:3}
\end{equation}\color{black}
when subjected to boundary conditions as per the following
\begin{equation}
	\boldsymbol{u}(\mathbf{y})\Big\vert_{\partial \Omega} = \tilde{\boldsymbol{u}}(\mathbf{y})\Big\vert_{\partial \Omega} + \bar{\boldsymbol{u}}(\mathbf{y})\Big\vert_{\partial \Omega}  \label{eq:4}
\end{equation}
Here, the displacement field mentioned earlier is decomposed into two parts. One part, \(\bar{\boldsymbol{u}}(\mathbf{y})\), represents the average displacement field, while \(\tilde{\boldsymbol{u}}(\mathbf{y})\) accounts for fluctuations due to the heterogeneous structure of the domain. The average displacement field \(\bar{\boldsymbol{u}}(\mathbf{y})\) is derived from the macroscale deformation field. To enforce the periodicity of this microscale volume element, \(\tilde{\boldsymbol{u}}(\mathbf{y})\) is required to follow the following relations:

\begin{eqnarray*}
	\tilde{\boldsymbol{u}}(\mathbf{y})\bigg\vert_{\mathbf{y}\in\partial\Omega^{ab}} &=& \tilde{\boldsymbol{u}}(\mathbf{y})\bigg\vert_{\mathbf{y}\in\partial\Omega^{dc}}\\
	\tilde{\boldsymbol{u}}(\mathbf{y})\bigg\vert_{\mathbf{y}\in\partial\Omega^{bc}} &=& \tilde{\boldsymbol{u}}(\mathbf{y})\bigg\vert_{\mathbf{y}\in\partial\Omega^{ad}}
\end{eqnarray*}

This micromechanical boundary value problem employs FE analysis for obtaining the field variables. Here, the micromechanical domain is divided into elements, and the solution is computed at integration points. The accuracy of this solution hinges on the level of discretization or the number of integration points utilized, directly impacting computational demands. To mitigate these costs, reduction of order methods can be applied and the domain is partitioned into $\mathcal{M}$ subdomains;

\begin{equation}
	\Omega=\bigcup_{i=1}^{\mathcal{M}}\Omega^{(i)} \label{eq:41}
\end{equation}
with each subdomain referring to a volume fraction $v_f^{(i)}$. 

The microscale field variable is assigned for each subdomain based on the distribution in the RVE.   
The field variable for each partition, $\bar{\boldsymbol{\beta}}^{(i)}$ is designated as 
\begin{equation}
	\bar{\boldsymbol{\beta}}^{(i)}=\int_{\Omega} \phi^{(i)} (\mathbf{y})\boldsymbol{\beta} (\mathbf{y})\,d\mathbf{y} \label{eq:nonlocal}
\end{equation}
where $\phi^{(i)} (\mathbf{y})$ is a weighing function. Furthermore, the following relation is utilized to obtain the distribution for the field quantity;
\begin{equation}
	\boldsymbol{\beta}(\mathbf{y})=\sum_{i=1}^\mathcal{M} \mathsf{N}^{(i)}(\mathbf{y})\bar{\boldsymbol{\beta}}^{(i)} \label{eq:reduced_order}
\end{equation}
The shape function, $\mathsf{N}^{(i)}(\mathbf{y})$ satisfies the partition of unity property i.e. $\sum_{i=1}^\mathcal{M} \mathsf{N}^{(i)}(\mathbf{y})=1$. Overall, the order of the RVE system reduces to $\mathcal{M}$. $\phi^{(i)} (\mathbf{y})$ ensures the nonlocal nature of the field variables and $\phi^{(i)} (\mathbf{y})$ should be positive ($\ge$0) with normality as $\displaystyle\int_{\Omega} \phi^{(i)} (\mathbf{y})d\mathbf{y}=1$.

Eq. (\ref{eq:reduced_order}) and (\ref{eq:nonlocal}) leads to condition of orthonormality of $\phi^{(i)} (\mathbf{y})$ with $\mathsf{N}^{(j)}(\mathbf{y})$ as
\begin{equation}
	\int_{\Omega} \phi^{(i)} (\mathbf{y})\mathsf{N}^{(j)}(\mathbf{y})d\mathbf{y}=\delta_{(ij)} \label{eq:44}
\end{equation}
Here $\delta_{(ij)}$ is the Kronecker delta with maximum indices values equal to the number of partitions. The choice of $\mathsf{N}^{(j)}(\mathbf{y})$ and $\phi^{(i)} (\mathbf{y})$ controls both the computational complexity and accuracy. In the proposed formulation, the following form of shape function and weighing function is taken:
\begin{align*}
	\mathsf{N}^{(i)}(\mathbf{y})=\begin{cases}
		1, & \text{if $\mathbf{y}\in \Omega^{(i)}$}\\
		0, & \text{if $\mathbf{y}\notin \Omega^{(i)}$}
	\end{cases}           & \hspace{14mm} \phi^{(i)} (\mathbf{y})=\dfrac{1}{\vert\Omega^{(i)}\vert}\mathsf{N}^{(i)}(\mathbf{y})
\end{align*}
Using this form of $\mathsf{N}^{(i)}(\mathbf{y})$ and $\phi^{(i)}(\mathbf{y})$ leads to the piecewise uniform distribution of the field for the partitions in microscale domain.

\section{Damage Informed Transformation Field Analysis (D-FEA)}\label{sec:3}
Transformation field analysis (TFA) employs the notion of transformation strain, also known as eigenstrain, which refers to a stress-free strain. Eigenstrain can be classified as "physical", such as residual strain, thermal strain, plastic strain, and others, as well as "fictitious" or "equivalent", such as damage-equivalent strain.

%
%
%
%
Here, for brevity, we focus on damage-equivalent eigenstrains and ignore all other kinds of eigenstrains. The damage in the material causes degradation of the material stiffness and alter the local fields caused by the externally applied mechanical loads. This modulation in the local strain fields can be accounted using equivalent eigenstrain field, $\boldsymbol{\mu}_d(\mathbf{y})$ impregnated in the homogeneous material. Alternatively, a eigenstress field, $\boldsymbol{\lambda}_d(\mathbf{y})$ adjusts the local stress fields to account the damage induced effects. Truly speaking, sometimes the equivalent eigenstrain field concept does not perfectly align with the physical eigenstrain field theory. The equivalent eigenstrain is fundamentally caused by the applied mechanical loads whereas the source of the physical eigenstrain is temperature or phase change, magnetic field or moisture concentration etc. Using same framework for both the cases is ambiguous. \color{black} In damage mechanics, particularly in micromechanics-based approaches or homogenization frameworks, damage can be associated with or cause an eigenstrain-like effect at the microscale. For instance, the local degradation of stiffness can lead to internal strain incompatibilities that are modeled as eigenstrains.\color{black}
 
First of all, the damage equivalent eigenstrain is defined as following
\begin{equation}
\boldsymbol{\mu}_d(\mathbf{y})=\mathbf{D}(\mathbf{y}):\boldsymbol{\epsilon}(\mathbf{y})
\end{equation}
where $\mathbf{D}(\mathbf{y})$ is damage tensor. Considering the isotropic damage state in all the phases, $\mathbf{D}(\mathbf{y})$ can be denoted in terms of a single damage variable, $\omega(\mathbf{y})$ and unit tensor, $\mathbf{I}$;
\begin{equation}
	\mathbf{D}(\mathbf{y})=\omega(\mathbf{y})\mathbf{I} \label{eq:12}
\end{equation}
 The resulting stress field can be represented using constitutive tensor, $\mathbf{L}(\mathbf{y})$ as 
\begin{equation}
\boldsymbol{\sigma}(\mathbf{y})=\mathbf{L}(\mathbf{y}):(\boldsymbol{\epsilon}(\mathbf{y})-\boldsymbol{\mu}_d(\mathbf{y})) \label{eq:11}
\end{equation}
Similarly, the damage equivalent eigenstress, $\boldsymbol{\lambda}_d(\mathbf{y})$ can be defined as following
\begin{equation}
\boldsymbol{\lambda}_d(\mathbf{y})=\mathbf{D}(\mathbf{\mathbf{y}}):\mathbf{L}(\mathbf{\mathbf{y}}):\boldsymbol{\epsilon}(\mathbf{\mathbf{y}})
\end{equation}
and corresponding the stress field is 
\begin{equation}
\boldsymbol{\sigma}(\mathbf{y})=\mathbf{L}(\mathbf{y}):\boldsymbol{\epsilon}(\mathbf{y})-\boldsymbol{\lambda}_d(\mathbf{y})
\end{equation}
Rate form of Eq. (\ref{eq:11}) is 
\begin{equation}
	\dot{\boldsymbol{\sigma}}(\mathbf{y})=[(\mathbf{I}-\mathbf{D}(\mathbf{y})):\mathbf{L}(\mathbf{y})]:\dot{\boldsymbol{\epsilon}}(\mathbf{y})-\dot{\omega}(\mathbf{y})\mathbf{I}:\mathbf{\color{black}L}\color{black}(\mathbf{y})\color{black}:\boldsymbol{\epsilon}(\mathbf{y}) \label{eq:15}
\end{equation} 
Eq. (\ref{eq:15}) shows that stress increment depends on the evolution of two state variables (a). strain    $\boldsymbol{\epsilon}$, and (b). damage, $\omega$. Furthermore, a dissipation potential function, $\mathfrak{F}^D$, accounting for the growth of damage, is considered as 
\begin{equation}
	\mathfrak{F}^D(Y(\mathbf{y}),\hspace{1mm} \omega(\mathbf{y})) = \dfrac{Y(\mathbf{y})}{(1-\omega(\mathbf{y}))}\left(\dfrac{\kappa_F}{\kappa_F-\kappa_D}\right)\dfrac{1}{\kappa_D} \label{eq:35}
\end{equation}
which gives the damage evolution as 
\begin{equation}
	\dot{\omega}(\mathbf{y})=\dot{\zeta}^D\dfrac{\partial \mathfrak{F}^D}{\partial Y}	\label{eq:36}
\end{equation}
where the damage growth force, $Y(\mathbf{y})$ and damage multiplier, $\dot{\zeta}^D$ are defined as 
\begin{eqnarray}
	Y(\mathbf{y})&=&\frac{1}{2}{\boldsymbol{\epsilon}}(\mathbf{y}):\mathbf{L}(\mathbf{y}):{\boldsymbol{\epsilon}}(\mathbf{y})\\
	\dot{\zeta}^D&=&(1-\omega(\mathbf{y}))\hspace{1mm}\dot{\kappa} \hspace{1mm}\boldsymbol{\mathsf{H}} (\kappa-\kappa_D)	 \label{eq:37}
\end{eqnarray}
$\boldsymbol{\mathsf{H}}(\cdot)$ is Heaviside step function and $\kappa$ is maximum principal strain. $\kappa_D$ is the threshold value of strain for initiation of damage, and $\kappa_F$ corresponds to strain for the state of complete failure/damage. 

Applying reduction of order technique and using Eq. (\ref{eq:nonlocal}) and Eq. (\ref{eq:reduced_order}), the microscopic state variables $\boldsymbol{\epsilon} (\mathbf{y})$, $\boldsymbol{\mu} (\mathbf{y})$ 
and ${\omega} (\mathbf{y})$ for RVE can be expressed in terms of weighing function, $\phi^{(i)} (\mathbf{y})$ as
\begin{align}
\bar{\boldsymbol{\epsilon}}^{(i)}=\int_{\Omega} \phi^{(i)} (\mathbf{y})\boldsymbol{\epsilon} (\mathbf{y})\,d\mathbf{y};          & \hspace{14mm} {\boldsymbol{\epsilon}}(\mathbf{y})=\sum_{i=1}^\mathcal{M} \mathsf{N}^{(i)}(\mathbf{y})\bar{{\boldsymbol{\epsilon}}}^{(i)}; \\
\bar{\boldsymbol{\mu}}^{(i)}=\int_{\Omega} \phi^{(i)} (\mathbf{y})\boldsymbol{\mu} (\mathbf{y})\,d\mathbf{y};          & \hspace{14mm} {\boldsymbol{\mu}}(\mathbf{y})=\sum_{i=1}^\mathcal{M} \mathsf{N}^{(i)}(\mathbf{y})\bar{{\boldsymbol{\mu}}}^{(i)}; \\
\bar{\omega}^{(i)}=\int_{\Omega} \phi^{(i)} (\mathbf{y})\omega (\mathbf{y})\,d\mathbf{y};          & \hspace{14mm} {\omega}(\mathbf{y})=\sum_{i=1}^\mathcal{M} \mathsf{N}^{(i)}(\mathbf{y})\bar{{\omega}}^{(i)}; \label{eq:omega}  
\end{align}
\subsection{Influence Tensors}
In TFA, a fourth-order tensor, $\mathbf{E}(\mathbf{y})$, which is called an elastic influence tensor, is defined to relate macroscopic strain, $\boldsymbol{\epsilon}^o$ with microscale distribution of the strain, $\boldsymbol{\epsilon}(\mathbf{y})$ as
\begin{equation}
	\boldsymbol{\epsilon}(\mathbf{y})=\mathbf{E}(\mathbf{y}):\boldsymbol{\epsilon}^o+\int_{\Omega}\mathbf{S}(\mathbf{y},\tilde{\mathbf{y}}):\boldsymbol{\mu}^o(\tilde{\mathbf{y}})\, d\tilde{\mathbf{y}}   \label{eq:49}
\end{equation}
$\mathbf{S}(\mathbf{y},\tilde{\mathbf{y}})$ is eigenstrain influence tensor function that relates the mechanical strain at a position $\mathbf{y}$ caused due to the eigenstrain $\boldsymbol{\mu}^o$ at another position $\tilde{\mathbf{y}}$. 
Finally, the Eq. (\ref{eq:49}) is expressed as
\begin{equation}
	\boldsymbol{\epsilon}(\mathbf{y})=\mathbf{E}(\mathbf{y}):\boldsymbol{\epsilon}^o+\sum_{j=1}^\mathcal{M}\left(\int_{\Omega} \mathsf{N}^{(j)}(\tilde{\mathbf{y}})\mathbf{S}(\mathbf{y},\tilde{\mathbf{y}}) \, d\tilde{\mathbf{y}}\right):\bar{\boldsymbol{\mu}}^{(j)}  \label{eq:52}
\end{equation}
replacing $\displaystyle\int_\Omega \mathsf{N}^{(j)}(\tilde{\mathbf{y}})\mathbf{S}(\mathbf{y},\tilde{\mathbf{y}}) \, d\tilde{\mathbf{y}}= \mathbf{S}^{(j)}(\mathbf{y})$ and multiplying both the sides by $\phi^{(i)} (\mathbf{y})$  and integrating it over \color{black}subdomain \color{black} $\Omega^i$ gives
\begin{equation}
	{\bar{\boldsymbol{\epsilon}}^{(i)}=\bar{\mathbf{E}}^{(i)}:\boldsymbol{\epsilon}^o+ \sum_{j=1}^\mathcal{M}\bar{\mathbf{S}}^{(ij)}:\bar{\boldsymbol{\mu}}^{(j)}} \label{eq:55}
\end{equation}
where $\bar{\mathbf{S}}^{(ij)}$ is eigen influence tensor, defined as
\begin{equation}
	\bar{\mathbf{S}}^{(ij)}=\int_{\Omega}\phi^{(i)} (\mathbf{y})\mathbf{S}^{(j)}(\mathbf{y})\, d\mathbf{y}   \label{eq:54}
\end{equation}
\color{black} TFA relies on precomputed influence functions (Green-type operators) that relate unit eigenstrain fields to resulting stress/strain fields. In this context, superposition of the eigenstrain contributions is a core principle of the method (as expressed by Eq. (\ref{eq:52})). This holds because of the linearity of the elastic Green's function used in influence tensor calculation. Additionally, the damage is modeled implicitly through the evolution of the eigenstrain field; not by reducing stiffness directly, but by growing eigenstrain magnitudes in damaged regions. However, this superposition is approximate; it assumes small perturbations, linear elastic responses to eigenstrains, and may not fully capture strong nonlinear interactions (e.g., severe localization or coalescing cracks).

\color{black}
A special partitioning scheme is adopted where the phase boundaries always coincide with the subvolume boundaries, and the existence of two different phases in a single subvolume is prohibited. Meanwhile, Eq. (\ref{eq:omega}) and Eq. (\ref{eq:12}) give 
\begin{align}
	\bar{\mathbf{D}}^{(i)}=\int_{\Omega} \phi^{(i)} (\mathbf{y})\mathbf{D} (\mathbf{y})\,d\mathbf{y};          & \hspace{14mm} {\mathbf{D}}(\mathbf{y})=\sum_{i=1}^\mathcal{M} \mathsf{N}^{(i)}(\mathbf{y})\bar{{\mathbf{D}}}^{(i)}; \label{eq:damage}
\end{align} 
However, the constitutive tensor, $\mathbf{L}(\mathbf{y})$ can be expressed as 
\begin{equation}
	\mathbf{L}(\mathbf{y})=(\mathbf{I}-\bar{\mathbf{D}}^{(i)}):\mathbf{L}^{\alpha(i)} \label{eq:58}
\end{equation}
Using Eq. (\ref{eq:damage}) and Eq. (\ref{eq:58}), the stress in a subvolume $i$ is calculated by multiplying $\boldsymbol{ \sigma}(\mathbf{y})=\mathbf{L}(\mathbf{y}):\boldsymbol{\epsilon}(\mathbf{y})$ with $ \phi^{(i)} (\mathbf{y})$ and taking the integration over the subvolume;
\begin{equation}
	\bar{\boldsymbol{\sigma}}^{(i)}=\left[(\mathbf{I}-\bar{\mathbf{D}}^{(i)}):\mathbf{L}^{\alpha(i)}:\bar{\boldsymbol{\epsilon}}^{(i)}\right] 
\end{equation}
where $\bar{\boldsymbol{\sigma}}^{(i)}=\displaystyle\int_{\Omega} \phi^{(i)} (\mathbf{y})\boldsymbol{\sigma} (\mathbf{y})\,d\mathbf{y}$ and $\alpha$ is a number assigned to phase material. Finally, using Eq. (\ref{eq:55}), stress in any partition $i$ is written as
\begin{equation}
	\bar{\boldsymbol{\sigma}}^{(i)}=(\mathbf{I}-\bar{\mathbf{D}}^{(i)}):\mathbf{L}^{\alpha(i)}:\left[\bar{\mathbf{E}}^{(i)}:\boldsymbol{\epsilon}^o+ \sum_{j=1}^\mathcal{M}\bar{\mathbf{S}}^{(ij)}:\bar{\boldsymbol{\mu}}^{(j)}\right]    \label{eq:60}
\end{equation}
\subsection{Macroscale Constitutive Response} \label{Macro_Constitutive}
As per the average stress theorem, the average stress in the RVE is equal to a constant stress tensor, $\boldsymbol{\sigma}^o$ with equivalent tractions, $\boldsymbol{t}^o\big\vert_{\partial\Omega}=\boldsymbol{\sigma}^o\boldsymbol{n}\big\vert_{\partial\Omega}$ prescribed over the domain boundary. A macroscale damage tensor $\bar{\mathbf{D}}$ is defined and the average stress in the RVE is expressed as;
\begin{equation}
	\bar{\boldsymbol{\sigma}}=(\mathbf{I}-\bar{\mathbf{D}}):\left[\dfrac{1}{\vert \Omega\vert}\int_\Omega\mathbf{L}(\mathbf{y}):\boldsymbol{\epsilon}(\mathbf{y})\,d\mathbf{y} \right]    \label{eq:62}
\end{equation}
substituting $\boldsymbol{\epsilon}(\mathbf{y})$ from Eq. (\ref{eq:52}), Eq. (\ref{eq:62}) is expressed as:
\begin{equation}
	\bar{\boldsymbol{\sigma}}=(\mathbf{I}-\bar{\mathbf{D}}):\left\{\dfrac{1}{\vert \Omega\vert}\int_\Omega \mathbf{L}(\mathbf{y}):\mathbf{E}(\mathbf{y})\,d\mathbf{y}\right\}:\boldsymbol{\epsilon}^o+(\mathbf{I}-\bar{\mathbf{D}}):\left\{\sum_{i=1}^\mathcal{M}\left(\dfrac{1}{\vert \Omega\vert}\int_\Omega \mathbf{L}(\mathbf{y}):\mathbf{S}^{(i)}(\mathbf{y})\,d\mathbf{y}\right)\right\}:\bar{\boldsymbol{\mu}}^{(i)} \label{eq:65}
\end{equation}
Eventually, it is represented as
\begin{equation}
	{\bar{\boldsymbol{\sigma}}=\bar{\mathbf{L}}_d:\boldsymbol{\epsilon}^o+\sum_{i=1}^\mathcal{M}\bar{\mathbf{M}}^{(i)}_d:\bar{\boldsymbol{\mu}}^{(i)}} \label{eq:66}
\end{equation}
These constitutive tensors, $\bar{\mathbf{L}}_d$ and $\bar{\mathbf{M}}^{(i)}_d$ can be concisely represented as  
\begin{eqnarray}
\bar{\mathbf{L}}_d&=&(\mathbf{I}-\bar{\mathbf{D}}):\bar{\mathbf{L}}\\
\bar{\mathbf{M}}^{(i)}_d&=&(\mathbf{I}-\bar{\mathbf{D}}):\bar{\mathbf{M}}_o^{(i)}
\end{eqnarray}
where 
\begin{eqnarray}
\bar{\mathbf{L}}&=&\left(\dfrac{1}{\vert \Omega \vert}\displaystyle\int_\Omega \mathbf{L}(\mathbf{y}):\mathbf{E}(\mathbf{y})\,d\mathbf{y}\right) \label{eq:L_bar} \\
\bar{\mathbf{M}}_o^{(i)}&=&\left(\dfrac{1}{\vert {\Omega}\vert}\displaystyle\int_\Omega \mathbf{L}(\mathbf{y}):\mathbf{S}^{(i)}(\mathbf{y})\,d\mathbf{y}\right) \label{eq:M_bar}
\end{eqnarray}

\subsection{Microscale and Macroscale Damage}
Microscale and macroscale damage can be derived by loading the RVE under uniform non-mechanical auxiliary stress $\Delta \tilde{\boldsymbol{\sigma}}$ or strain $\Delta \tilde{\boldsymbol{\epsilon}}$ fields. Here, '$\Delta$' signifies the strain state with reference to another threshold value corresponding to initiation of damage in the material. Non-mechanical nature of applied field signifies the absence of any boundary forces/displacements or body forces.  Let us consider the local eigenstrain, $\boldsymbol{\mu}(\mathbf{y})$ due to applied auxiliary strain field which is particularly a damage induced eigenstrain caused due to distributed damage $\mathbf{D}(\mathbf{y})$ in the RVE. Furthermore, the damage induced eigenstrain for each \color{black}subdomain \color{black} is expressed as 
\begin{equation}
\bar{{\boldsymbol{\mu}}}^{(i)}=\bar{{\mathbf{D}}}^{(i)}:\Delta \tilde{\boldsymbol{\epsilon}}
\end{equation}
Similarly, the local eigenstress, $\boldsymbol{\lambda}(\mathbf{y})$ due to applied auxiliary stress field causes the damage-induced eigen stress field for each \color{black}subdomain \color{black} as 
\begin{equation}
\bar{{\boldsymbol{\lambda}}}^{(i)}=\bar{{\mathbf{D}}}^{(i)}:\Delta \tilde{\boldsymbol{\sigma}}
\end{equation}
The macroscopic eigenstrain and eigenstress are pursued due to the uniform auxiliary strain and stress fields as 
\begin{equation}
\bar{{\boldsymbol{\mu}}}=\bar{{\mathbf{D}}}:\Delta \tilde{\boldsymbol{\epsilon}}
\end{equation}
\begin{equation}
\bar{{\boldsymbol{\lambda}}}=\bar{{\mathbf{D}}}:\Delta \tilde{\boldsymbol{\sigma}}
\end{equation}
For damage, different loading scenarios give rise to two methods to account eigenstrain or eigenstress for obtaining the response of the material as explained in the next sections.

\subsubsection{Uniform strain field condition}
Uniform strain field condition corresponds to the state of RVE under auxiliary strain field $\Delta \tilde{\boldsymbol{\epsilon}}$ and overall macroscopic damage $\bar{{\mathbf{D}}}_{\epsilon}$ caused by the auxiliary strain field is calculated by considering the two different loading conditions. First loading condition refers to eigenstrain fields $\bar{{\boldsymbol{\mu}}}^{(i)}$ (sub-domains) and $\bar{{\boldsymbol{\mu}}}$ (overall) caused by the auxiliary strain field $\Delta \tilde{\boldsymbol{\epsilon}}$.  Macroscopic and kinematically admissible microscopic fields are
\begin{align*}
	\mathsf{Macro} & \hspace{15mm}^{[\bf{1}]}\bar{\boldsymbol{\sigma}} = 0  & \hspace{5mm} \\
	& \hspace{15mm} ^{[\bf{1}]}\bar{{\boldsymbol{\epsilon}}}=\bar{{\mathbf{D}}}_{\epsilon}:\Delta \tilde{\boldsymbol{\epsilon}}  \\
	\mathsf{Micro} & \hspace{15mm} {^{[\bf{1}]}\boldsymbol{\epsilon}}^{(i)}(\mathbf{y}) = \mathbf{M}^{(i)}:{^{[\bf{1}]}\boldsymbol{\sigma}}^{(i)}(\mathbf{y}) +  {\bar{{\mathbf{D}}}_{\epsilon}}^{(i)}:\Delta \tilde{\boldsymbol{\epsilon}} 
\end{align*}
For second loading condition, there exists a uniform stress state $^{[\bf{2}]}\bar{\boldsymbol{\sigma}}={\boldsymbol{\sigma}}^{\circ}$ in the RVE and leads to overall and statically admissible local fields as following
\begin{align*}
	\mathsf{Macro} & \hspace{15mm}^{[\bf{2}]}\bar{\boldsymbol{\sigma}} = {\boldsymbol{\sigma}}^{\circ}   & \hspace{5mm} \\
	\mathsf{Micro} & \hspace{15mm} ^{[\bf{2}]}{\boldsymbol{\sigma}}^{(i)}(\mathbf{y}) = \mathbf{B}^{(i)}(\mathbf{y}):^{[\bf{2}]}\bar{\boldsymbol{\sigma}} \\
	& \hspace{15mm} ^{[\bf{2}]}{\boldsymbol{\epsilon}}^{(i)}(\mathbf{y}) = \mathbf{M}^{(i)}:^{[\bf{2}]}{\boldsymbol{\sigma}}^{(i)}(\mathbf{y})  
\end{align*}
The virtual work equation for kinematically admissible strain field of loading state '1', and statically admissible stress field of loading state '2', is 
\begin{equation}
	\dfrac{1}{\vert\Omega\vert} \int_{\Omega}  \{{^{[\bf{1}]}\boldsymbol{\sigma}}(\mathbf{y})\}:\{{^{[\bf{2}]}\boldsymbol{\epsilon}}^{(i)}(\mathbf{y})\}\,d\mathbf{y}=^{[\bf{1}]}\bar{\boldsymbol{\sigma}}:^{[\bf{2}]}\bar{\boldsymbol{\epsilon}}  =0
\end{equation}
or
\begin{equation}
	\int_{\Omega}  \{{^{[\bf{1}]}\boldsymbol{\sigma}}^{(i)}(\mathbf{y})\}:\mathbf{M}^{(i)}:\{{^{[\bf{2}]}\boldsymbol{\sigma}}^{(i)}(\mathbf{y})\}\, d\mathbf{y}=0
\end{equation}
since ${^{[\bf{2}]}\boldsymbol{\sigma}}^{(i)}(\mathbf{y})\neq 0$, it applies that
\begin{equation}
	{^{[\bf{1}]}\boldsymbol{\sigma}}^{(i)}(\mathbf{y})=0
\end{equation}
Another virtual work equation for other two fields is
\begin{equation}
	\dfrac{1}{\vert\Omega\vert}\int_{\Omega}  \{{^{[\bf{2}]}\boldsymbol{\sigma}}^{(i)}(\mathbf{y})\}:\{{^{[\bf{1}]}\boldsymbol{\epsilon}}^{(i)}(\mathbf{y})\}\,d\mathbf{y}=^{[\bf{2}]}\bar{\boldsymbol{\sigma}}:^{[\bf{1}]}\bar{\boldsymbol{\epsilon}} 
\end{equation} 
which can be expressed as 
\begin{equation}
	\dfrac{1}{\vert\Omega\vert}\int_{\Omega}  \mathbf{B}^{(i)}(\mathbf{y}):^{[\bf{2}]}\bar{\boldsymbol{\sigma}}: {\bar{{\mathbf{D}}}_{\epsilon}}^{(i)}:\Delta \tilde{\boldsymbol{\epsilon}}\,d\mathbf{y}=^{[\bf{2}]}\bar{\boldsymbol{\sigma}}:\bar{{\mathbf{D}}}_{\epsilon}:\Delta \tilde{\boldsymbol{\epsilon}} 
\end{equation}
which finally gives
\begin{equation}
	\bar{{\mathbf{D}}}_{\epsilon}=\dfrac{1}{\vert\Omega\vert}\int_{\Omega}  \mathbf{B}^{\text{T}(i)}(\mathbf{y}): {\bar{{\mathbf{D}}}_{\epsilon}}^{(i)}\,d\mathbf{y}=\sum_{i=1}^\mathcal{M} v_f^{(i)} \bar{\mathbf{B}}^{\text{T}(i)}:{{\bar{\mathbf{D}}}_{\epsilon}}^{(i)}
\end{equation} 
\subsubsection{Uniform stress field condition}
Uniform stress field condition caused by the auxiliary stress  $\Delta \tilde{\boldsymbol{\sigma}}$ leads macroscopic damage state, $\bar{{\mathbf{D}}}_{\sigma}$ in the RVE.  Again two loading conditions are considered where one loading condition refers to eigenstress field $\bar{{\boldsymbol{\lambda}}}^{(i)}$ (sub-domains) and $\bar{{\boldsymbol{\lambda}}}$ (overall) caused by the auxiliary stress field $\Delta \tilde{\boldsymbol{\sigma}}$. Macroscopic and statically admissible microscale fields are 
\begin{align*}
	\mathsf{Macro} & \hspace{15mm}^{[\bf{1}]}\bar{\boldsymbol{\epsilon}} = 0  & \hspace{5mm} \\
	& \hspace{15mm} ^{[\bf{1}]}\bar{{\boldsymbol{\sigma}}}=\bar{{\mathbf{D}}}_{\sigma}:\Delta \tilde{\boldsymbol{\sigma}}  \\
	\mathsf{Micro} & \hspace{15mm} ^{[\bf{1}]}\bar{\boldsymbol{\sigma}}^{(i)}(\mathbf{y}) = \mathbf{L}^{(i)}:^{[\bf{1}]}\bar{\boldsymbol{\epsilon}}^{(i)}(\mathbf{y}) +  {\bar{{\mathbf{D}}}_{\sigma}}^{(i)}:\Delta \tilde{\boldsymbol{\sigma}} 
\end{align*}
Another loading conditions corresponding to uniform strain state $^{[\bf{2}]}\bar{\boldsymbol{\epsilon}}={\boldsymbol{\epsilon}}^{\circ}$ is sought which yields overall and statically admissible local fields as following
\begin{align*}
	\mathsf{Macro} & \hspace{15mm}^{[\bf{2}]}\bar{\boldsymbol{\epsilon}} = {\boldsymbol{\epsilon}}^{\circ}   & \hspace{5mm} \\
	\mathsf{Micro} & \hspace{15mm} ^{[\bf{2}]}{\boldsymbol{\epsilon}}^{(i)}(\mathbf{y})= \mathbf{A}^{(i)}(\mathbf{y}):^{[\bf{2}]}\bar{\boldsymbol{\epsilon}} \\
	& \hspace{15mm} ^{[\bf{2}]}{\boldsymbol{\sigma}}^{(i)}(\mathbf{y}) = \mathbf{L}^{(i)}:{^{[\bf{2}]}\boldsymbol{\epsilon}}^{(i)}(\mathbf{y})
\end{align*}   
The virtual work equation for kinematically admissible strain field of loading state '1', and statically admissible stress field of loading state '2', is 
\begin{equation}
	\dfrac{1}{\vert\Omega\vert} \int_{\Omega}  \{{^{[\bf{2}]}\boldsymbol{\sigma}}^{(i)}(\mathbf{y})\}:\{{^{[\bf{1}]}\boldsymbol{\epsilon}}^{(i)}(\mathbf{y})\}\,d\mathbf{y}=^{[\bf{2}]}\bar{\boldsymbol{\sigma}}:^{[\bf{1}]}\bar{\boldsymbol{\epsilon}}  =0
\end{equation}
or
\begin{equation}
	\int_{\Omega}  \mathbf{L}^{(i)}:{^{[\bf{2}]}\boldsymbol{\epsilon}}^{(i)}(\mathbf{y}):{^{[\bf{1}]}\boldsymbol{\epsilon}}^{(i)}(\mathbf{y})\,d\mathbf{y}=0
\end{equation}
since $\{{^{[\bf{2}]}\boldsymbol{\epsilon}}^{(i)}(\mathbf{y})\}\neq 0$, it applies that
\begin{equation}
	{^{[\bf{1}]}\boldsymbol{\epsilon}}^{(i)}(\mathbf{y})=0
\end{equation}
Another virtual work equation for other two fields is
\begin{equation}
	\dfrac{1}{\vert\Omega\vert}\int_{\Omega}  \{{^{[\bf{1}]}\boldsymbol{\sigma}}^{(i)}(\mathbf{y})\}:\{{^{[\bf{2}]}\boldsymbol{\epsilon}}^{(i)}(\mathbf{y})\}\,d\mathbf{y}=^{[\bf{1}]}\bar{\boldsymbol{\sigma}}:^{[\bf{2}]}\bar{\boldsymbol{\epsilon}}
\end{equation} 
which can be expressed as 
\begin{equation}
	\dfrac{1}{\vert\Omega\vert}\int_{\Omega}  { \bar{{\mathbf{D}}}_{\sigma}}^{(i)}:\Delta \tilde{\boldsymbol{\sigma}}:\mathbf{A}^{(i)}(\mathbf{y}):^{[\bf{2}]}\bar{\boldsymbol{\epsilon}}\,d\mathbf{y}=\bar{{\mathbf{D}}}_{\sigma}:\Delta \tilde{\boldsymbol{\sigma}}: ^{[\bf{2}]}\bar{\boldsymbol{\epsilon}}
\end{equation}
which finally gives
\begin{equation}
	\bar{{\mathbf{D}}}_{\sigma}=\dfrac{1}{\vert\Omega\vert}\int_{\Omega}  \mathbf{A}^{\text{T}(i)}(\mathbf{y}):{ \bar{{\mathbf{D}}}_{\sigma}}^{(i)}\,d\mathbf{y}=\sum_{i=1}^\mathcal{M} v_f^{(i)} \bar{\mathbf{A}}^{\text{T}(i)}:{{\bar{\mathbf{D}}}_{\sigma}}^{(i)}
\end{equation} 
The above mentioned expression holds for a heterogeneous representative volume when subjected to uniform auxiliary stress.

Under generalized multiaxial loading conditions, the damage tensor is calculated using a factor $\psi$ to account the effect of both the fields; 
\begin{equation}
		\bar{{\mathbf{D}}}=\psi \bar{{\mathbf{D}}}_{\epsilon}+(1-\psi) \bar{{\mathbf{D}}}_{\sigma}
\end{equation}
where $\psi \in \{0,1\}$ signifies the contribution of uniform strain state in the overall damage evaluation. 

For fiber reinforced composites, the loading along the fibers constitutes the uniform strain condition where as transverse to fibers loading embodies the uniform stress field. When the load is applied along the fiber direction, the material fails due to breakage of fibers, however, the matrix damage happens due to transverse direction loading.  Considering this the material state can determined using $\psi$ as    	
\begin{equation}
		\psi = \left(\frac{\mathfrak{R}_{f}}{\mathfrak{R}_{f}+\mathfrak{R}_{m}}\right)
\end{equation}	

Hence the damage in fiber direction and transverse to fiber direction can be evaluated using factors $\mathfrak{R}_{f}$ and $\mathfrak{R}_{m}$ which are adopted from Hashin's criteria \cite{hashin1963variational} for failure of 3D fiber reinforced composites.  

Failure in fiber direction:
		\begin{equation}
			\mathfrak{R}_{f}=\left(\frac{\epsilon_{ 11}}{\mathfrak{E}_{11}}\right)^{2}+\left(\frac{\epsilon_{ 12}}{\mathfrak{E}_{12}}\right)^{2}+\left(\frac{\epsilon_{ 13}}{\mathfrak{E}_{13}}\right)^{2} 
		\end{equation}
		
Failure in transverse direction:
		\begin{equation}
			\mathfrak{R}_{m}=\left(\frac{\epsilon_{22}}{\mathfrak{E}_{22}}\right)^{2}+\left(\frac{\epsilon_{33}}{\mathfrak{E}_{22}}\right)^{2}+\left(\frac{\epsilon_{23}^{2}-\epsilon_{ 22} \epsilon_{ 33}}{\mathfrak{E}_{23}^{2}}\right)+\left(\frac{\epsilon_{12}^{2}+\epsilon_{ 13}^{2}}{\mathfrak{E}_{12}^{2}}\right) 
		\end{equation}
where $\mathfrak{E}_{11}$ and $\mathfrak{E}_{22}$ are the strain values corresponding to tensile strengths in the fiber and transverse directions respectively, $\mathfrak{E}_{13}$, $\mathfrak{E}_{23}$, $\mathfrak{E}_{12}$ represent shear strains at shear strengths along out of plane and in-plane directions of ply. Here, the condition $\mathfrak{R}_{f}\geqslant 1$ signifies the state of damage initiation along fiber direction and likewise, the $\mathfrak{R}_{m}\geqslant 1$ corresponds the onset of damage in transverse direction. These expressions demonstrate the contributions of directional strains in fiber and matrix failure.
\color{black}
\section{Clustering Driven Reduced Order Modelling} \label{sec:4}

\color{black}In TFA-based methods, the RVE domain is systematically discretized into numerous smaller subdomains, often referred to as partitions. Each subdomain is assigned a unique transformation field variable that encapsulates the representative behavior of that region. Traditionally, this partitioning is often performed heuristically or based on user intuition, without a systematic or response-driven strategy. As a result, the partitioning can lead to two extremes: a coarse discretization (with fewer partitions) that yields lower accuracy but faster computation, or a fine discretization (with many partitions) that improves accuracy at the cost of higher computational expense. To address this challenge, we introduce a $k$-means clustering-based partitioning scheme that utilizes the local mechanical response of the RVE to generate an optimized partitioning map. \color{black}

In past, a data-driven, cluster-based Reduced Order Model (ROM) was proposed for computational homogenization of heterogeneous materials, aimed at reducing computational complexity. The proposed method leverages a clustering technique, originally introduced by \citep{Liu2016, Liu2018a, Liu2018b}, to group finite elements within the RVE based on the similarity of their response tensors. Building upon the success of data-driven clustering, FCA and specially CNTFA in predicting heterogeneous material properties (\cite{Kafka2018, Liu2018b, Li2019, Shakoor2019, Yu2019, cheng2019fem, nie2019principle, nie2021efficient, ri2021cluster}), we adapt this method to proposed TFA based multiscale framework. A critical step involves determining suitable loading conditions for computing the strain tensor, which serves as input for data-driven clustering. To predict failure morphologies in RVEs using eigen-strain driven TFA, we apply three or six load vectors with periodic boundary conditions to 2D or 3D RVEs, respectively. The resulting strain tensors, obtained from FEM simulations, are used for clustering.

For each integration point of the RVE domain, a microscale field variable is evaluated using predefined set of loading. Furthermore, based on the response of the RVE partitioning map is constructed in such a way so that all the elements in a single partition have nearly uniform response. This response has been checked either in terms of 1). elastic strain tensor ${\boldsymbol{\epsilon}}(\mathbf{y})$, or 2). eigen strain tensor ${\boldsymbol{\mu}}(\mathbf{y})$. Selection of elastic strain tensor is inspired from CNTFA where as, since the failure in RVE is essentially \color{black}governed \color{black} through the damage-equivalent eigen strains, eigen strain tensor is picked. $k$-means clustering scheme is adopted to group the elements into partitions. Based on the response function, two types of $k$-means clustering schemes are investigated as:
\begin{enumerate}
	\item Elastic Clustering 
	\item Eigen Clustering
\end{enumerate}  
In the first category of $k$-means clustering scheme, the elements are grouped into $k$ number of clusters according to the elastic/total strain components ($\epsilon_{11}$, $\epsilon_{22}$, $\epsilon_{33}$, $\epsilon_{23}$, $\epsilon_{13}$ and $\epsilon_{12}$). These components are obtained from six loading conditions which are imposed over the boundary of the RVE:
\begin{eqnarray}
		\mathfrak{L}_1: \hspace{5mm}\boldsymbol{\epsilon}^o &=& \upvartheta \boldsymbol{e}_1 \otimes \boldsymbol{e}_1 \label{cl_eq1}\\
		\mathfrak{L}_2: \hspace{5mm}\boldsymbol{\epsilon}^o &=& \upvartheta \boldsymbol{e}_2 \otimes \boldsymbol{e}_2 \label{cl_eq2} \\
		\mathfrak{L}_3: \hspace{5mm}\boldsymbol{\epsilon}^o &=& \upvartheta \boldsymbol{e}_3 \otimes \boldsymbol{e}_3 \label{cl_eq3}\\
		\mathfrak{L}_4: \hspace{5mm}\boldsymbol{\epsilon}^o &=& \dfrac{\upvartheta}{2} \boldsymbol{e}_1 \otimes \boldsymbol{e}_2 + \dfrac{\upvartheta}{2} \boldsymbol{e}_2 \otimes \boldsymbol{e}_1 \label{cl_eq4}\\
		\mathfrak{L}_5: \hspace{5mm}\boldsymbol{\epsilon}^o &=& \dfrac{\upvartheta}{2} \boldsymbol{e}_2 \otimes \boldsymbol{e}_3 + \dfrac{\upvartheta}{2} \boldsymbol{e}_3 \otimes \boldsymbol{e}_2 \label{cl_eq5}\\
		\mathfrak{L}_6: \hspace{5mm}\boldsymbol{\epsilon}^o &=& \dfrac{\upvartheta}{2} \boldsymbol{e}_3 \otimes \boldsymbol{e}_1 + \dfrac{\upvartheta}{2} \boldsymbol{e}_1 \otimes \boldsymbol{e}_3 \label{cl_eq6}
\end{eqnarray}
The value of $\upvartheta$ is selected arbitrarily to consider only the elastic effects without any damage in the domain.

For eigen clustering, in order to capture the damage-induced eigenstrain, $\upvartheta$ is increased till the damage initiates in RVE. The damage-induced eigenstrains are calculated at all the integration points as 
\begin{equation}
	\boldsymbol{{\mu}}=\omega\mathbf{I}:\boldsymbol{\epsilon}
\end{equation}  
Further, the damage-induced eigenstrain components ($\mu_{11}$, $\mu_{22}$, $\mu_{33}$, $\mu_{23}$, $\mu_{13}$ and $\mu_{12}$) are used for clustering the elements. Let us denote $\mathcal{H}(\mathbf{y})$ a vector consisting of all the components of elastic strain or eigenstrain obtained from all the loading cases, $\mathfrak{L}_1$, $\mathfrak{L}_2$.....$\mathfrak{L}_6$ which is defined as
\begin{eqnarray} \footnotesize
	\mathcal{H}(\mathbf{y})=\begin{cases}
		\left\{\underbrace{\epsilon^{\mathfrak{L}_1}_{11}(\mathbf{y}),\epsilon^{\mathfrak{L}_1}_{22}(\mathbf{y}),......,\epsilon^{\mathfrak{L}_1}_{12}(\mathbf{y}),......}_{\mathfrak{L}_1},\underbrace{\epsilon^{\mathfrak{L}_2}_{11}(\mathbf{y}),\epsilon^{\mathfrak{L}_2}_{22}(\mathbf{y}),......,\epsilon^{\mathfrak{L}_2}_{12}(\mathbf{y}),.......}_{\mathfrak{L}_2},......\right\}^{\text{T}}_{36\times 1} & \text{$\mathsf{Elastic \hspace{1mm}Clustering}$}\vspace{3mm} \\
		\left\{\underbrace{\mu^{\mathfrak{L}_1}_{11}(\mathbf{y}),\mu^{\mathfrak{L}_1}_{22}(\mathbf{y}),.....,\mu^{\mathfrak{L}_1}_{12}(\mathbf{y}),.....}_{\mathfrak{L}_1},\underbrace{\mu^{\mathfrak{L}_2}_{11}(\mathbf{y}),\mu^{\mathfrak{L}_2}_{22}(\mathbf{y}),.....,\mu^{\mathfrak{L}_2}_{12}(\mathbf{y}),......}_{\mathfrak{L}_2},......\right\}^{\text{T}}_{36\times 1} & \text{$\mathsf{Eigen \hspace{1mm}Clustering}$} \label{eq:h_vector}
	\end{cases}
\end{eqnarray} 
Furthermore, the Euclidean norm is calculated by considering two integration points $\mathbf{y}_1$ and $\mathbf{y}_2$, as  
\begin{equation}
	\left\Vert  \mathcal{H}(\mathbf{y}_1)-\mathcal{H}(\mathbf{y}_2) \right\Vert_2 = \sqrt{\sum_{i=1}^{36}\left(\mathcal{H}_i(\mathbf{y}_1)-\mathcal{H}_i(\mathbf{y}_2)\right)^2 }
\end{equation}
The $k$-means clustering scheme aims to partition $n$ integration points into $k$ number of clusters such that each integration point belongs to a cluster with minimum distance from its corresponding cluster center or nearest mean. Let us take I$^{th}$ cluster from a set of all clusters and $\bar{\mathcal{H}}_I$ is the mean of $\mathcal{H}(\mathbf{y})$ residing in I$^{th}$ cluster. The objective is to minimize the sum of distance between $\bar{\mathcal{H}}_I$ and other $\mathcal{H}(\mathbf{y}_i)$ where $i$ belongs to same cluster 'I'. The clusters are designated as $\Omega^{(1)}$, $\Omega^{(2)}$, $\Omega^{(3)}$,....,$\Omega^{({\mathcal{M}})}$.
\begin{equation}
	\mathcal{S}=\underset{\mathcal{S'}}{\arg\min}\sum_{I=1}^{\mathcal{M}}\sum_{i\in \Omega_I}\left\Vert  \mathcal{H}(\mathbf{y}_i)-\bar{\mathcal{H}}_I \right\Vert_2 \label{eq:dist_fun}
\end{equation}    
Minimization of the cumulative distance function results a particular clustering map $\mathcal{S}=$($\Omega^{(1)}$, $\Omega^{(2)}$, $\Omega^{(3)}$,....,$\Omega^{({\mathcal{M}})}$). Not necessarily, the adjacent integration points lie in the same partition and ultimately the procedure may results the scattered distribution of single cluster.

\section{Computational Implementation Procedure}\label{Sec_NumericalP}\label{sec:5}

\begin{figure}[H]
	\centering
	\includegraphics[width=1\textwidth]{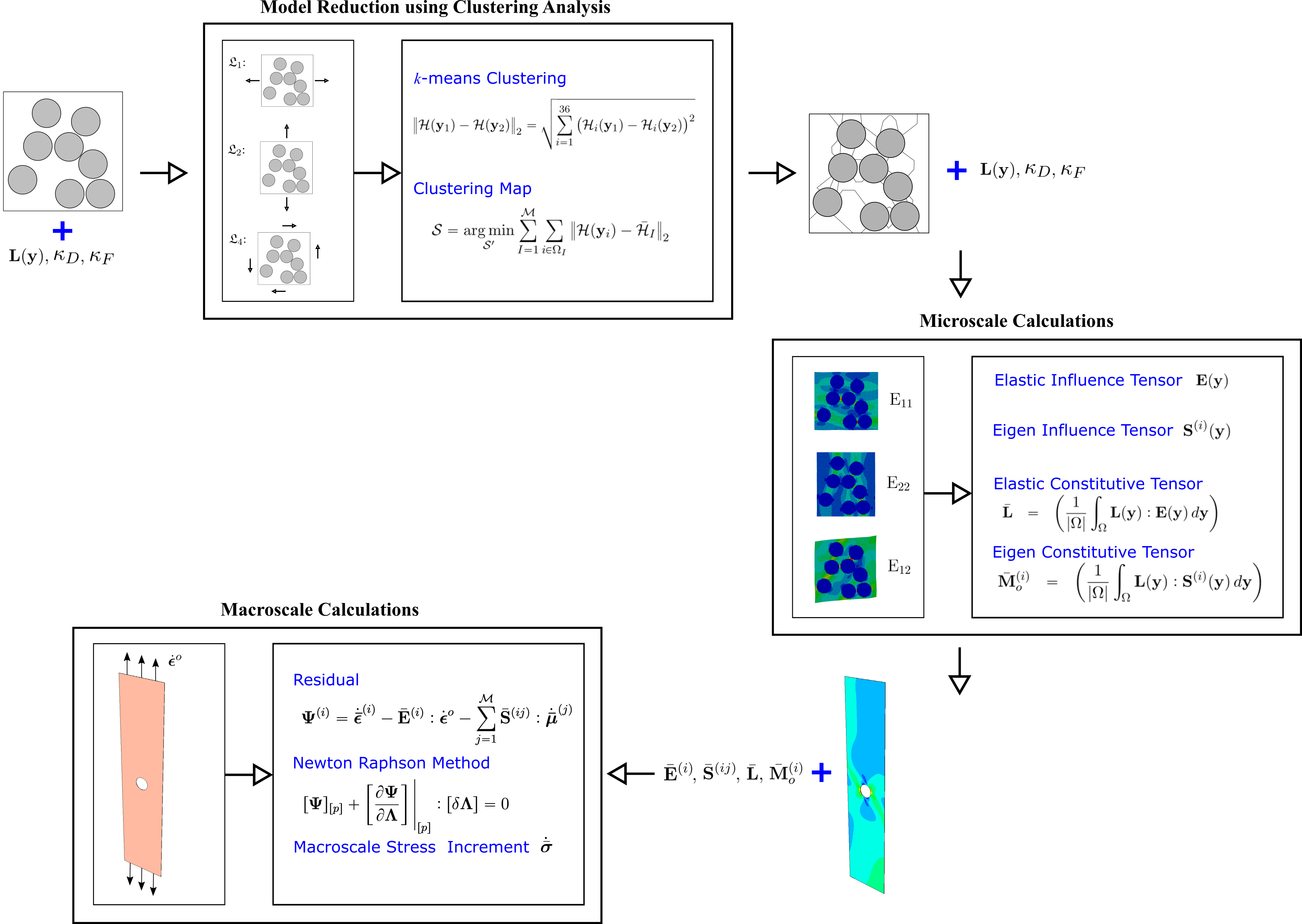}
	\caption{The numerical procedure for the implementation of multiscale formulation begins with defining the microstructural domain and reducing its order through clustering analysis. Once partitioning is completed, influence and homogenized property tensors are calculated. Later, the macroscale boundary value problem is formulated. Field variables are then determined by solving nonlinear equations using the Newton-Raphson algorithm and the previously calculated influence and homogenized property tensors.}
	\label{fig:method}
\end{figure}

The proposed formulation is implemented through a two-stage numerical procedure. The first stage, referred to as the offline stage, begins with defining the microstructural domain and performing order reduction via clustering analysis. Once the domain is partitioned, the influence tensors and homogenized property tensors are computed. The second stage, known as the online stage, involves formulating the macroscale boundary value problem. Subsequently, the field variables are determined by solving the nonlinear equations using the Newton-Raphson algorithm, along with the influence and homogenized property tensors as shown in Fig. \ref{fig:method}. The following sections provide the steps for performing the microscale and macroscale calculations. 

\subsection{Offline Stage - I: Model Reduction using Clustering Analysis}
This stage begins by characterizing the RVE through definition of its morphology, which corresponds to a fixed volume fraction of heterogeneities. The configuration of an RVE includes number and position of heterogenities corrresponding to either a volume fraction or size of heterogenity. Furthermore, the model reduction procedure consists of following steps:
\begin{itemize}
\item The RVE domain is discretised into elements for FE analysis. Three configurations of RVE with 1). Single 2). Eight and 3). Thirty fibers have been \color{black}utilized \color{black} in this manuscript. All the three configurations correspond to same volume volume fraction. 
\item The material properties, in terms of $\mathbf{L}(\mathbf{y})$, $\kappa_D$ and $\kappa_F$ are assigned for each phase.
\item The boundary conditions are applied depending upon the loading cases to be simulated for different strain states in the RVE. A 3D RVE requires all six loading conditions as per Eq. (\ref{cl_eq1}) to Eq. (\ref{cl_eq6}) to be applied, however Eq. (\ref{cl_eq1}), (\ref{cl_eq2}) and (\ref{cl_eq4}) can be employed for obtaining the plain strain state in the RVE domain. 
 \item Next, we calculate the elastic strain ${\boldsymbol{\epsilon}}(\mathbf{y})$ and damage-induced eigen strain ${\boldsymbol{\mu}}(\mathbf{y})$ distribution in the RVE against the applied loading conditions. 
\item All the strain components for the elements are rearranged for the calculation of response vector, $\mathcal{H}(\mathbf{y})$ as per Eq. (\ref{eq:h_vector}).  
\item Finally the cumulative distance function $\mathcal{S}$ is calculated for fixed number of clusters, $\mathcal{M}$ as per Eq. (\ref{eq:dist_fun})) which results a particular clustering map $\mathcal{S}=\Omega^{(1)}$, $\Omega^{(2)}$, $\Omega^{(3)}$,....,$\Omega^{({\mathcal{M}})}$).
\end{itemize}
\color{black}

Our primary objective is to devise a multiscale framework for capturing the microscopic damage effects. Consequently, the RVEs used are artificially designed to illustrate microstructural features. It is crucial to note that determining RVEs (or Statistical Volume Elements) in actual random materials is a separate and complex undertaking that necessitates careful analysis (\cite{singh2023representative}).

\color{black}
 
\begin{figure}[H]
	\centering
	\includegraphics[width=1.0\textwidth]{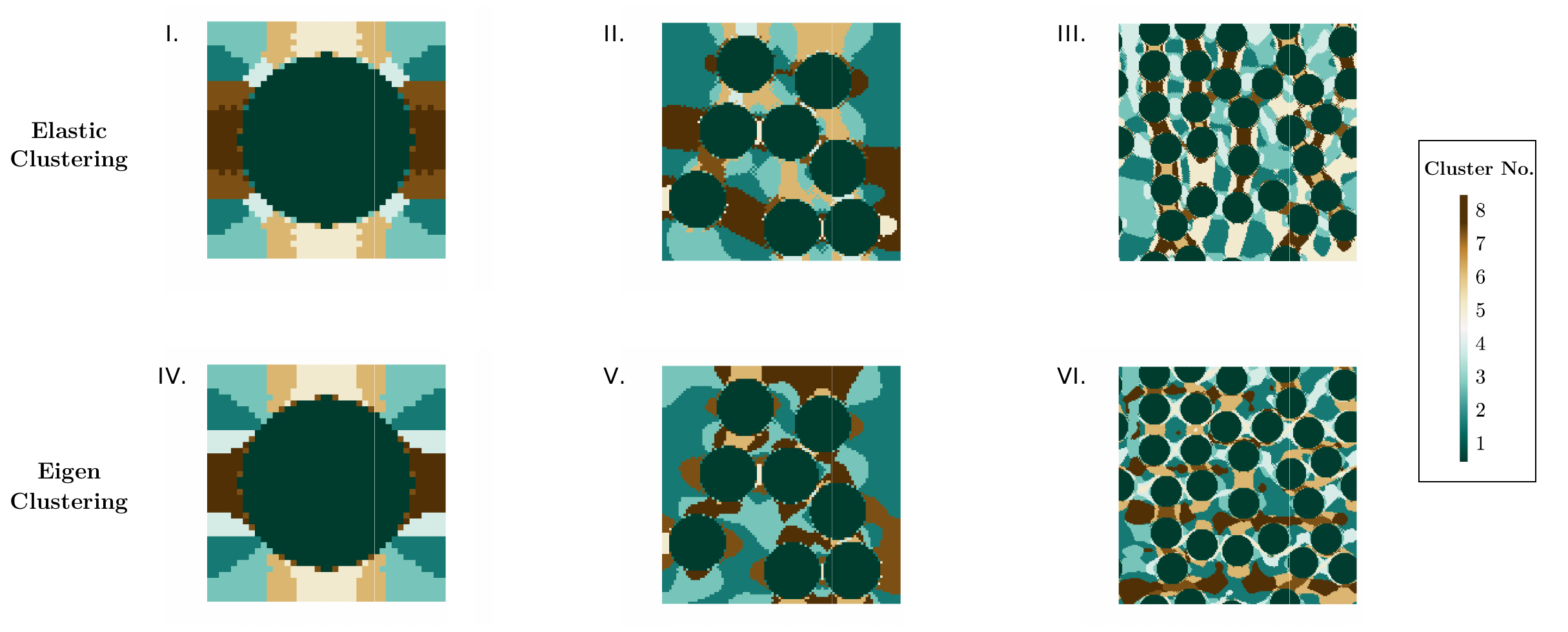}
	\caption{Two sets of clustering maps for three RVEs with one, eight, and thirty fibers, respectively. The first set ( I, II, and III) shows eight clusters ($\mathcal{M}=8$) obtained via elastic clustering. The second set (IV, V, and VI) is derived from eigen clustering.}
	\label{fig:2}
\end{figure}
The above mentioned clustering methodology is applied for three RVE configurations consisting different number of randomly placed fibers. Fig. \ref{fig:2} depicts two set of clustering maps for three RVEs with single, eight and thirty number of fibers, respectively. First set (Fig. \ref{fig:2} I, \ref{fig:2} II and \ref{fig:2} III) illustrates eight number of clusters ($\mathcal{M}=8$) obtained using the elastic clustering and second set (Fig. \ref{fig:2} IV, \ref{fig:2} V and \ref{fig:2} VI) is acquired through eigen clustering scheme. \color{black} Here, we demonstrate that for analysing any macroscale domain with heterogeneities, fracture analysis of an RVE is needed to determine the clustering map. Certainly, there are other methods available for the calculation of eigenstrain distribution in the microdomain. Instead of calculating the damage-induced eigenstrains, the thermal strains can also be calculated by applying the unit temperature change. \color{black}  
\begin{algorithm}\footnotesize
	\caption{Microscale Problem}\label{micro-problem}
	$\blacksquare$ \Input{\begin{itemize}
			\item[-] Characterization of RVE corresponding to a fixed volume fraction of heterogeneities 
			\vspace{-0.1cm}\item[-] Material Data: $\mathbf{L}(\mathbf{y})$, $\kappa_D$ and $\kappa_F$  for each phase
		\end{itemize}
	}\vspace{0.1cm}
	$\blacksquare$ \Output{\begin{itemize}
				\vspace{-0.1cm}\item[-] Partitioned RVE domain and each subdomain designated with a partition number, $i$ varies from 1 to $\mathcal{M}$.
			  	\vspace{-0.1cm}\item[-] Elastic Influence Tensor $\mathbf{E}(\mathbf{y})$ and $\bar{\mathbf{E}}$ 
			  	\vspace{-0.1cm}\item[-] Eigen Influence Tensor $\bar{\mathbf{S}}^{(ij)}$ 
			  	\vspace{-0.1cm}\item[-] Homogenized Elastic Constitutive Tensor $\bar{\mathbf{L}}$
			  	\vspace{-0.1cm}\item[-]  Homogenized Eigen Constitutive Tensor $\bar{\mathbf{M}}^{(i)}_o$
		\end{itemize}}
	\vspace{0.1cm}$\blacksquare$ \Begin(\hspace{2mm} \textbf{Model Reduction using Clustering Analysis}\vspace{2mm}){
		 Construct FE model for a given RVE.  \\ \vspace{0.1cm}
		 Assign elastic and damage material properties to each phase.\\ \vspace{0.1cm}
		 Apply all six loading conditions as per Eq. (\ref{cl_eq1}) to Eq. (\ref{cl_eq6}) for 3D RVE domain or Eq. (\ref{cl_eq1}), (\ref{cl_eq2}) and (\ref{cl_eq4}) for 2D microstructure.\\ \vspace{0.1cm} \hfill\Comment{\textcolor{black}{In the presence of periodic boundary conditions}}\\ \vspace{0.1cm}
		  Obtain the elastic strain ${\boldsymbol{\epsilon}}(\mathbf{y})$ and damage-induced eigen strain ${\boldsymbol{\mu}}(\mathbf{y})$ distribution in the RVE against the applied loading conditions. \\ \vspace{0.1cm}
		 Calculate of response vector, $\mathcal{H}(\mathbf{y})$ as per Eq. (\ref{eq:h_vector}).  \\ \vspace{0.1cm}
		 Calculate cumulative distance function $\mathcal{S}$ for fixed number of clusters, $\mathcal{M}$ as per Eq. (\ref{eq:dist_fun}).\\ \vspace{0.1cm}
		 Obtain the clustering map $\mathcal{S}=$($\Omega^{(1)}$, $\Omega^{(2)}$, $\Omega^{(3)}$,....,$\Omega^{({\mathcal{M}})}$).
		}
	\vspace{0.1cm}$\blacksquare$ \Begin(\hspace{2mm} \textbf{Influence and Homogenized Constitutive Tensors Calculations}\vspace{2mm}){
Initialize the calculations using partitioned RVE domain and label the subdomains with identifiers.\\ \vspace{0.1cm}  \hfill\Comment{\textcolor{black}{FE model of clustering analysis can be used}}\\ \vspace{0.1cm}
Assign elastic properties to each partition/cluster.\\ \vspace{0.1cm}
Apply six loading conditions for 3D RVE domain or three loading conditions for 2D microstructure for simulating for different strain states.\\ \vspace{0.1cm}
Calculate Elastic Influence Tensor $\mathbf{E}^{(i)}(\mathbf{y})$ and $\bar{\mathbf{E}}^{(i)}$ as per Eq. (\ref{eq:49}) and Eq. (\ref{eq:55}).\\ \vspace{0.1cm}
Calculate Eigen Influence Tensor $\bar{\mathbf{S}}^{(ij)}$ using $\bar{\mathbf{E}}^{(i)}$.\\ \vspace{0.1cm}
Calculate Homogenized Elastic Constitutive Tensor $\bar{\mathbf{L}}$ using Eq. (\ref{eq:L_bar}).\\ \vspace{0.1cm}
Calculate Homogenized Eigen Constitutive Tensor $\bar{\mathbf{M}}^{(i)}_o$ using Eq. (\ref{eq:M_bar}).\\ \vspace{0.1cm}
}		
\end{algorithm}

\subsection{Offline Stage - II: Influence and Homogenized Constitutive Tensors Calculations} 
Second phase of offline stage starts with the partitioned RVE domain obtained as an output of clustering analysis. Following are the steps for calculations of influence and homogenized constitutive tensors:
\begin{itemize}
	\item The first step is the designation of sub-domains with identifiers in FE model of microstructure domain. 
	\item Furthermore, elastic influence tensor $\mathbf{E}(\mathbf{y})$ or $\bar{\mathbf{E}}^{(i)}$ are calculated by performing FE analyses corresponding to six sets of macroscale strain components as per Eq. (\ref{eq:49}) and Eq. (\ref{eq:55}).
	\item Eigenstrain influence tensors, $\bar{\mathbf{S}}^{(ij)}$ are computed using the elastic influence tensors as given in Eq. (\ref{eq:54}). 
	\item The homogenized properties of RVE in terms of elastic constitutive tensor $\bar{\mathbf{L}}$ is evaluated using $\mathbf{E}(\mathbf{y})$ and microconstituents' material properties $\mathbf{L}(\mathbf{y})$ as per Eq. (\ref{eq:L_bar}).
	\item Secondly, eigen constitutive tensor, $\bar{\mathbf{M}}^{(i)}_o$ is calculated using $\mathbf{S}^{(i)}(\mathbf{y})$ and microconstituents' material properties $\mathbf{L}(\mathbf{y})$ as per Eq. (\ref{eq:M_bar}).
\end{itemize}
Algorithm-\ref{micro-problem} illustrates the procedure followed in the offline stage for solving the problem at the microscale.
    
\subsection{Online Stage: Macroscale Calculations}\label{sec:method}
Incremental forms of transformation field and constitutive equations are formulated for determining the macroscopic solution. This starts by rewriting Eq. (\ref{eq:55}) in incremental form as
\begin{equation}
	\dot{\boldsymbol{\bar{\epsilon}}}^{(i)}-\bar{\mathbf{E}}^{(i)}:\dot{\boldsymbol{{\epsilon}}}^o- \sum_{j=1}^\mathcal{M}\bar{\mathbf{S}}^{(ij)}:\dot{\boldsymbol{\bar{\mu}}}^{(j)}=0 \label{eq:77}
\end{equation}

Eq. (\ref{eq:77}) contains two unknowns 1). $\dot{\boldsymbol{\bar{\epsilon}}}^{(i)}$, and 2). $\dot{\boldsymbol{\bar{\mu}}}^{(j)}$. For calculating these a Newton-Raphson method is applied and a residual $\mathbf{\color{black}\Psi}^{(i)}$ is defined as
\begin{equation}
	\mathbf{\color{black}\Psi}^{(i)}=\dot{\boldsymbol{\bar{\epsilon}}}^{(i)}-\bar{\mathbf{E}}^{(i)}:\dot{\boldsymbol{{\epsilon}}}^o- \sum_{j=1}^\mathcal{M}\bar{\mathbf{S}}^{(ij)}:\dot{\boldsymbol{\bar{\mu}}}^{(j)} \label{eq:78}
\end{equation}
To solve this nonlinear system of equations for $[\mathbf{\color{black}\Psi}]=0$ in an iterative manner, gives:
\begin{equation}
	[\mathbf{\color{black}\Psi}]_{[p]}+\left[\dfrac{\partial\mathbf{\color{black}\Psi}}{\partial\mathbf{\color{black}\Lambda}}\right]\Bigg\vert_{[p]}:[\delta \mathbf{\color{black}\Lambda}]=0  \label{eq:82}
\end{equation}
where $[\mathbf{\color{black}\Psi}]=\Big\{\mathbf{\color{black}\Psi}^{(1)} \quad \mathbf{\color{black}\Psi}^{(2)} \quad \mathbf{\color{black}\Psi}^{(3)} \cdots  \mathbf{\color{black}\Psi}^{(\mathcal{M})}\Big\}$ represents a matrix consists of residual vectors and $[\mathbf{\color{black}\Lambda}]=\Big\{\mathbf{\color{black}\Lambda}^{(1)} \quad \mathbf{\color{black}\Lambda}^{(2)} \quad \mathbf{\color{black}\Lambda}^{(3)} \cdots  \mathbf{\color{black}\Lambda}^{(\mathcal{M})}\Big\}$ is another matrix defining $\mathbf{\color{black}\Lambda}^{(i)}=\dot{\boldsymbol{\bar{\epsilon}}}^{(i)}$. In Eq. (\ref{eq:82}), $p$ denotes the number of iterations required to calculate the final solution. Tangent matrix, $\left[\dfrac{\partial\mathbf{\color{black}\Psi}}{\partial\mathbf{\color{black}\Lambda}}\right]$ is expanded as 
\begin{equation}
	\left[\dfrac{\partial\mathbf{\color{black}\Psi}}{\partial\mathbf{\color{black}\Lambda}}\right]=
	\left[\renewcommand\arraystretch{2.0}
	\begin{matrix}
		\dfrac{\partial\mathbf{\color{black}\Psi}^{(1)}}{\partial\mathbf{\color{black}\Lambda}^{(1)}} & \dfrac{\partial\mathbf{\color{black}\Psi}^{(1)}}{\partial\mathbf{\color{black}\Lambda}^{(2)}} & \cdots & \dfrac{\partial\mathbf{\color{black}\Psi}^{(1)}}{\partial\mathbf{\color{black}\Lambda}^{(\mathcal{M}-1)}} &\dfrac{\partial\mathbf{\color{black}\Psi}^{(1)}}{\partial\mathbf{\color{black}\Lambda}^{(\mathcal{M})}} \\
		\dfrac{\partial\mathbf{\color{black}\Psi}^{(2)}}{\partial\mathbf{\color{black}\Lambda}^{(1)}} & \dfrac{\partial\mathbf{\color{black}\Psi}^{(2)}}{\partial\mathbf{\color{black}\Lambda}^{(2)}} & \cdots & \dfrac{\partial\mathbf{\color{black}\Psi}^{{(2)}}}{\partial\mathbf{\color{black}\Lambda}^{(\mathcal{M}-1)}} & \dfrac{\partial\mathbf{\color{black}\Psi}^{(2)}}{\partial\mathbf{\color{black}\Lambda}^{(\mathcal{M})}} \\
		\vdots & \vdots & \ddots & \vdots  & \vdots \\
		\dfrac{\partial\mathbf{\color{black}\Psi}^{(\mathcal{M}-1)}}{\partial\mathbf{\color{black}\Lambda}^{(1)}} & \dfrac{\partial\mathbf{\color{black}\Psi}^{(\mathcal{M}-1)}}{\partial\mathbf{\color{black}\Lambda}^{(2)}} & \cdots & \dfrac{\partial\mathbf{\color{black}\Psi}^{(\mathcal{M}-1)}}{\partial\mathbf{\color{black}\Lambda}^{(\mathcal{M}-1)}} & \dfrac{\partial\mathbf{\color{black}\Psi}^{(\mathcal{M}-1)}}{\partial\mathbf{\color{black}\Lambda}^{(\mathcal{M})}} \\
		\dfrac{\partial\mathbf{\color{black}\Psi}^{(\mathcal{M})}}{\partial\mathbf{\color{black}\Lambda}^{(1)}} & \dfrac{\partial\mathbf{\color{black}\Psi}^{(\mathcal{M})}}{\partial\mathbf{\color{black}\Lambda}^{(2)}} & \cdots & \dfrac{\partial\mathbf{\color{black}\Psi}^{(\mathcal{M})}}{\partial\mathbf{\color{black}\Lambda}^{(\mathcal{M}-1)}} & \dfrac{\partial\mathbf{\color{black}\Psi}^{(\mathcal{M})}}{\partial\mathbf{\color{black}\Lambda}^{(\mathcal{M})}}
	\end{matrix}
	\right]  \label{eq:83}
\end{equation}
which is expressed as
\begin{equation}
	\dfrac{\partial\mathbf{\color{black}\Psi}^{(i)}}{\partial\mathbf{\color{black}\Lambda}^{(j)}}=\mathbf{I}\delta_{ij}-\bar{\mathbf{E}}^{(i)}:\left(\dfrac{\partial\boldsymbol{\dot{\epsilon}}^o}{\partial\dot{\boldsymbol{\bar{\epsilon}}}^{(j)}}\right)- \sum_{k=1}^\mathcal{M}\bar{\mathbf{S}}^{(ik)}:\left(\dfrac{\partial\dot{\boldsymbol{\bar{\mu}}}^{(k)}}{\partial\dot{\boldsymbol{\bar{\epsilon}}}^{(j)}}\right) \label{eq:84}
\end{equation}
\begin{algorithm}\footnotesize
	\caption{Macroscale Problem}\label{macro-problem}
	$\blacksquare$ \Input{\begin{itemize}
	  	\item[-] Macroscale Strain Increment $\dot{\boldsymbol{{\epsilon}}}^o_{(n+1),[p]}$ 
	  	\vspace{-0.1cm}\item[-] Influence Tensors and Constitutive Tensors $\bar{\mathbf{E}}$, $\bar{\mathbf{S}}^{(ij)}$, $\bar{\mathbf{L}}$, $\bar{\mathbf{M}}_o^{(i)}$
	  	\vspace{-0.1cm}\item[-] Material Data: Microscopic - $\mathbf{L}(\mathbf{y})$, $\kappa^{(i)}_D$ and $\kappa^{(i)}_F$
	  	\vspace{-0.1cm}\item[-] Material Data: Macroscopic - $\mathfrak{E}_{11}$,  $\mathfrak{E}_{22}$, $\mathfrak{E}_{13}$, $\mathfrak{E}_{23}$, $\mathfrak{E}_{12}$
	  \end{itemize}
	}\vspace{-0.1cm}
	 $\blacksquare$ \Output{\begin{itemize}
	 		\vspace{-0.1cm}\item[-] Macroscopic Stress ${\bar{\boldsymbol{\sigma}}}_{(n+1),[p]}$\end{itemize}}
 \vspace{-0.1cm}$\blacksquare$ \Begin{ Initialize  $\dot{\boldsymbol{\bar{\mu}}}^{(j)}_{(n+1),[p]} \gets  \dot{\boldsymbol{\bar{\mu}}}^{(j)}_{(n),[p]}$  \\
	Initialize  $\dot{\boldsymbol{\bar{\epsilon}}}^{(i)}_{(n+1),[p]} \gets \big(\bar{\mathbf{E}}^{(i)}:\dot{\boldsymbol{{\epsilon}}}^o_{(n+1),[p]}- \sum_{j=1}^\mathcal{M}\bar{\mathbf{S}}^{(ij)}:\boldsymbol{\dot{\bar{\mu}}}^{(j)}_{(n+1),[p]}\big)$\\
	\For{$p\leftarrow 1$ \KwTo maximum iterations}
	{
		Update ${\boldsymbol{\bar{\epsilon}}}^{(i)}_{(n+1),[p]}={\boldsymbol{\bar{\epsilon}}}^{(i)}_{(n),[p]}+\dot{\boldsymbol{\bar{\epsilon}}}^{(i)}_{(n+1),[p]}$\\ 
	 Calculate ${\mathfrak{R}_{f}}_{(n+1),[p]}$, ${\mathfrak{R}_{m}}_{(n+1),[p]}$ and $\psi_{(n+1),[p]}$\\	 
	 Calculate ${{}\bar{{\mathbf{D}}}_{\epsilon}}^{(i)}_{(n+1),[p]}$ and ${{}\bar{{\mathbf{D}}}_{\sigma}}^{(i)}_{(n+1),[p]}$  using ${\boldsymbol{\bar{\epsilon}}}^{(i)}_{(n+1),[p]}$ \\
	 Calculate 		${{}\bar{{\mathbf{D}}}}^{(i)}_{(n+1),[p]}=\psi_{(n+1),[p]} {{}\bar{{\mathbf{D}}}_{\epsilon}}^{(i)}_{(n+1),[p]}+(1-\psi_{(n+1),[p]}) {{}\bar{{\mathbf{D}}}_{\sigma}}^{(i)}_{(n+1),[p]}$\\
	 Update ${\dot{{}\bar{{\mathbf{D}}}}}^{(i)}_{(n+1),[p]}={{}\bar{{\mathbf{D}}}}^{(i)}_{(n+1),[p]}-{{}\bar{{\mathbf{D}}}}^{(i)}_{(n),[p]}$\\
	 Calculate $\dot{\boldsymbol{\bar{\mu}}}^{(i)}_{(n+1),[p]}$ using ${\dot{{}\bar{{\mathbf{D}}}}}^{(i)}_{(n+1),[p]}$ and $\dot{\boldsymbol{\bar{\epsilon}}}^{(i)}_{(n+1),[p]}$\\
	 Calculate $\left(\dfrac{\partial\dot{\boldsymbol{{\epsilon}}}^o}{\partial\dot{\boldsymbol{\bar{\epsilon}}}^{(i)}}\right)_{(n+1),[p]}$ and $\left(\dfrac{\partial\dot{\boldsymbol{\bar{\mu}}}^{(i)}}{\partial\dot{\boldsymbol{\bar{\epsilon}}}^{(i)}}\right)_{(n+1),[p]}$ as per Eqs. (\ref{eq:98}) and (\ref{eq:89})\\
	 Calculate $\left[\dfrac{\partial\mathbf{\color{black}\Psi}^{(i)}}{\partial\mathbf{\color{black}\Lambda}^{(j)}}\right]_{(n+1),[p]}$ or $\left[\dfrac{\partial\mathbf{\color{black}\Psi}}{\partial\mathbf{\color{black}\Lambda}}\right]_{(n+1),[p]}$\\
	 Evaluate $\mathbf{\color{black}\Lambda}_{(n+1),[p+1]}=\mathbf{\color{black}\Lambda}_{(n+1),[p]}-\left[\dfrac{\partial\mathbf{\color{black}\Psi}}{\partial\mathbf{\color{black}\Lambda}}\right]^{-1}_{(n+1),[p]}:\mathbf{\color{black}\Psi}_{(n+1),[p]}$\\
	 Calculate $\left\Vert\mathbf{\color{black}\Lambda}_{(n+1),[p+1]}-\mathbf{\color{black}\Lambda}_{(n+1),[p]}\right\Vert_2$
	 
	 \eIf {$\left\Vert\mathbf{\color{black}\Lambda}_{(n+1),[p+1]}-\mathbf{\color{black}\Lambda}_{(n+1),[p]}\right\Vert_2>{TOL}$}
	  {$p \gets p+1$\\
	  $\dot{\boldsymbol{\bar{\epsilon}}}^{(i)}_{(n+1),[p+1]} \gets \mathbf{\color{black}\Lambda}_{(n+1),[p+1]}$}	 
	 {${\boldsymbol{\bar{\epsilon}}}^{(i)}_{(n+1),[p]}\gets{\boldsymbol{\bar{\epsilon}}}^{(i)}_{(n),[p]}+\dot{\boldsymbol{\bar{\epsilon}}}^{(i)}_{(n+1),[p]}$\\${\boldsymbol{\bar{\mu}}}^{(i)}_{(n+1),[p]}\gets{\boldsymbol{\bar{\mu}}}^{(i)}_{(n),[p]}+\dot{\boldsymbol{\bar{\mu}}}^{(i)}_{(n+1),[p]}$}
	 {Stop}
	 }
	 Calculate $\bar{\mathbf{L}}_d=(\mathbf{I}-{{}\bar{{\mathbf{D}}}}_{(n+1),[p]}):\bar{\mathbf{L}}$ and 
	 $\bar{\mathbf{M}}^{(i)}_d=(\mathbf{I}-{{}\bar{{\mathbf{D}}}}^{(i)}_{(n+1),[p]}):\bar{\mathbf{M}}_o^{(i)}$\\
	 Calculate Macroscopic Stress ${\bar{\boldsymbol{\sigma}}_{(n+1),[p]}=\bar{\mathbf{L}}_d:\boldsymbol{\epsilon}^o_{(n+1),[p]}+\sum_{i=1}^\mathcal{M}\bar{\mathbf{M}}^{(i)}_d:{\boldsymbol{\bar{\mu}}}^{(i)}_{(n+1),[p]}}$}
\end{algorithm}

Furthermore, the derivative of eigenstrain increment with respect to total strain increment is expressed as 
\begin{equation}
	\left(\dfrac{\partial\dot{\boldsymbol{\bar{\mu}}}^{(i)}}{\partial\dot{\boldsymbol{\bar{\epsilon}}}^{(i)}}\right)=\mathbf{I}-({{\mathbf{L}}^{\alpha{(i)}}})^{-1}:\left(\dfrac{\partial\boldsymbol{\dot{\bar{\sigma}}}^{(i)}}{\partial\dot{\boldsymbol{\bar{\epsilon}}}^{(i)}}\right) \label{eq:89}
\end{equation}
where using Eq. (\ref{eq:11}) and (\ref{eq:15}) tangent stiffness, $\left(\dfrac{\partial\dot{\boldsymbol{\bar{\sigma}}}^{(i)}}{\partial\dot{\boldsymbol{\bar{\epsilon}}}^{(i)}}\right)$ can be expressed as:
\begin{equation}
	\left(\dfrac{\partial\dot{\boldsymbol{\bar{\sigma}}}^{(i)}}{\partial\dot{\boldsymbol{\bar{\epsilon}}}^{(i)}}\right)=\left({{\mathbf{L}}^{\alpha{(i)}}}(\bar{\mathbf{D}}^{(i)})+\frac{\partial{{{\mathbf{L}}^{\alpha{(i)}}}(\bar{\mathbf{D}}^{(i)})}}{\partial{\bar{\boldsymbol{\epsilon}}^{(i)}}}:\boldsymbol{\dot{\bar{\epsilon}}}^{(i)}\right)  \label{eq:91}
\end{equation}
and
\begin{equation}
	\frac{\partial{{{\mathbf{L}}^{\alpha{(i)}}}(\bar{\mathbf{D}}^{(i)})}}{\partial{\bar{\boldsymbol{\epsilon}}^{(i)}}}=\left(\dfrac{\partial\mathbf{L}^{\alpha{(i)}}(\bar{\mathbf{D}}^{(i)})}{\partial\bar{\mathbf{D}}^{(i)}}\right):\left(\dfrac{\partial\bar{\mathbf{D}}^{(i)}}{\partial\omega^{(i)}}\right)\times\left(\dfrac{\partial\omega^{(i)}}{\partial{\kappa^{(i)}}}\right):\left(\dfrac{\partial\kappa^{(i)}}{\partial{\bar{\boldsymbol{\epsilon}}^{(i)}}}\right)  \label{eq:92}
 \end{equation}
However, we also have
\begin{equation}
	\left(\dfrac{\partial\dot{\boldsymbol{{\epsilon}}}^o}{\partial\dot{\boldsymbol{\bar{\epsilon}}}^{(i)}}\right)={(\bar{\mathbf{E}}^{(i)})}^{-1}:\left[\mathbf{I}-\sum_{j=1}^\mathcal{M}\bar{\mathbf{S}}^{(ij)}:\left(\dfrac{\partial\dot{\boldsymbol{\bar{\mu}}}^{(j)}}{\partial\dot{\boldsymbol{\bar{\epsilon}}}^{(j)}}\right)\delta_{ij}\right] \label{eq:98}
\end{equation}
The steps for applying the Newton Raphson scheme for solving the macroscale problem is discussed in Algorithm-\ref{macro-problem}.

\section{Computational Verification}\label{sec:6}
Fig. \ref{fig:method} outlines the sequential implementation of the D-TFA based homogenization procedure, encompassing two key stages: preprocessing and macroscale solution. This methodology is implemented within a finite element method framework. A two-tiered verification process is conducted. Firstly, RVE simulations are performed under both tensile and shear loading conditions utilizing the elastic and eigen clustering driven D-TFA approach. Subsequently, multi-scale analyses are conducted for 3-dimensional heterogeneous domains subjected to complex loading conditions to validate the code implementation. The results of macroscale analysis are then compared with experimental data available in the literature. Table \ref{table:0} provides a summary of the diverse numerical studies undertaken in this work to showcase the capabilities of the proposed methodology. 

\vspace{0.75cm}
\begin{table}[h]
\captionsetup{width=0.75\textwidth}
    \caption{\color{black}Numerical studies conducted with the D-TFA technique}
 \centering
\begin{tabular}{c p{5.75cm} c p{7cm}}
\toprule
\textbf{No.} & \textbf{Name of the study} & \textbf{Domain} & \textbf{Purpose}\\
\midrule
1 & RVE simulations - I & 2D & To determine the effect of clustering for single, eight and thirty fibers RVE.\\
2 & RVE simulations - II & 2D & To compare of elastic and eigen clustering schemes.\\
3 & RVE simulations - III & 2D & To estimate of microscale damage morphology and compare with FEM. \\
4 & RVE simulations - IV & 3D & To perform tensile and shear characterization under six different loading scanerios.\\
5 & Open hole specimen under tension & 3D & To investigate the laminate response under tensile load for various open hole specimens with different fiber orientations and predict the crack trajectories. \\
\bottomrule
\end{tabular} \label{table:0}
\end{table}
\subsection{Numerical Implementation-1: Determination of effect of clustering for single, eight and thirty fibers RVE}

Microscale analysis is carried out for three two-dimensional microscale domains 1). single fiber RVE, 2). eight fiber RVE and 3). thirty fiber RVE. \color{black} We utilize the Random Sequential Adsorption (RSA) technique, a widely adopted method for generating microstructures with randomly placed heterogeneities. In RSA, fibers are placed one by one at random positions. If a new fiber overlaps with an existing one, it is rejected, and a new random position is tried. A MATLAB code is written to implement RSA for obtaining fiber center coordinates, aiming to create a specific number of non-overlapping circular fibers within a 1-unit square RVE. Fiber radii are determined by the desired fiber volume fraction and the number of fibers. All three cases corresponds to fiber volume fraction of 41\%. 

Geometric periodicity, which ensures the continuity of fibers at RVE boundaries, is implemented by mimicking the unit cell at its edges so that while placing a particle, if its center is near a boundary of the RVE, it must also "wrap around" to the opposite side. \color{black} Geometric periodicity is maintained by introducing the same fibers on the opposite boundaries and Periodic boundary conditions are enforced in all directions except for the direction of the applied load. The RVE is modeled with a finite element mesh containing 49284 plane strain quadrilateral elements. The microstructural RVE used in this set of numerical examples represents a two-phase material system, specifically a composite comprising fibers embedded in matrix. The phase properties of RVEs are summarized in Table \ref{table:RVE_properties}. \color{black}The post damage response of composite under transverse loading depends significantly on the matrix material's damage parameters i.e. damage initiation strain and damage failure strain. For these simulations, we used the damage failure strain as approximately four times the damage initiation strain, resulting in the gradual decrease of post-damage stress with strain, this allowing us to demonstrate the effectiveness of proposed $k$-means-based domain partitioning approach within the computational framework. \color{black}

\vspace{0.75cm}
\begin{table}[h]
	\captionsetup{width=0.75\textwidth}
	\caption{Material properties of phase materials used for RVE simulations. Source: \cite{singh2017reduced}}
	
	\centering
	\begin{tabular}{lcc}
		\toprule
		\textbf{Material Property} & \textbf{Fiber} & \textbf{Matrix} \\
		\midrule
		Volume Fraction & 0.41 & 0.59\\
		Elastic Modulus [MPa]& 80,000 & 2,670 \\
		Poisson's Ratio & 0.3 & 0.3 \\
		Damage Initiation Strain [\%] & - & 0.9 \\
		Damage Failure Strain [\%] & - & 3.15 \\
		\bottomrule
	\end{tabular} \label{table:RVE_properties}
\end{table}
\vspace{0.5cm}  

\begin{figure}[H]
\centering
\includegraphics[height=0.925\textheight]{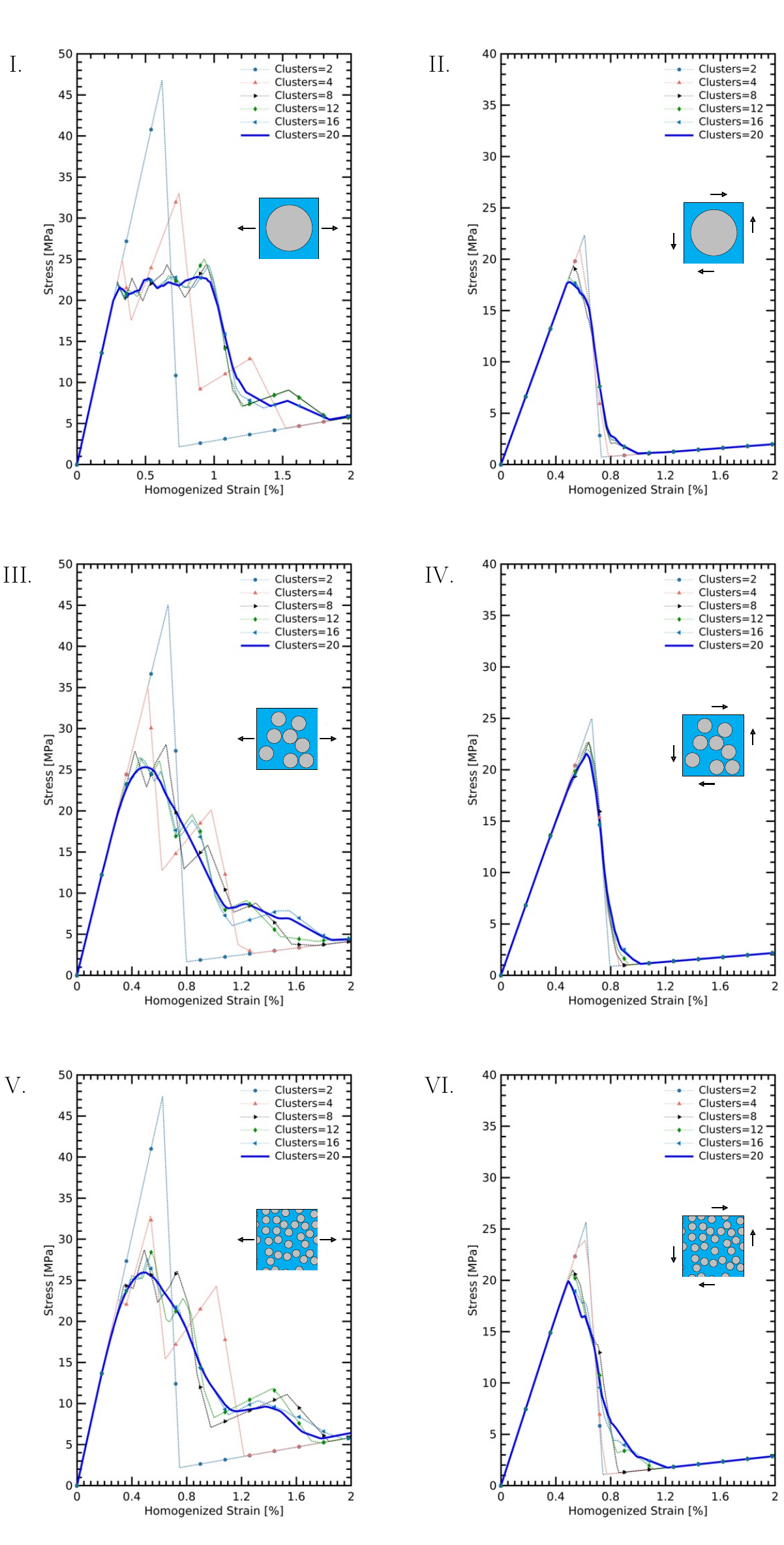}
\caption{Homogenized stress-strain response of single-fiber RVE (subfigures I and II), eight-fiber RVE (subfigures III and IV) and thirty-fiber RVE (subfigures V and VI) when subjected to tensile (subfigures I, III and V) and shear loading (subfigures II, IV and VI). The response is checked for different number of clusters ($\mathcal{M}=$ 2, 4, 8, 12, 16 and 20) obtaining by performing \textbf{elastic clustering scheme}}
\label{fig:elastic_cluster}
\end{figure}

\begin{figure}[H]
\centering
\includegraphics[height=0.925\textheight]{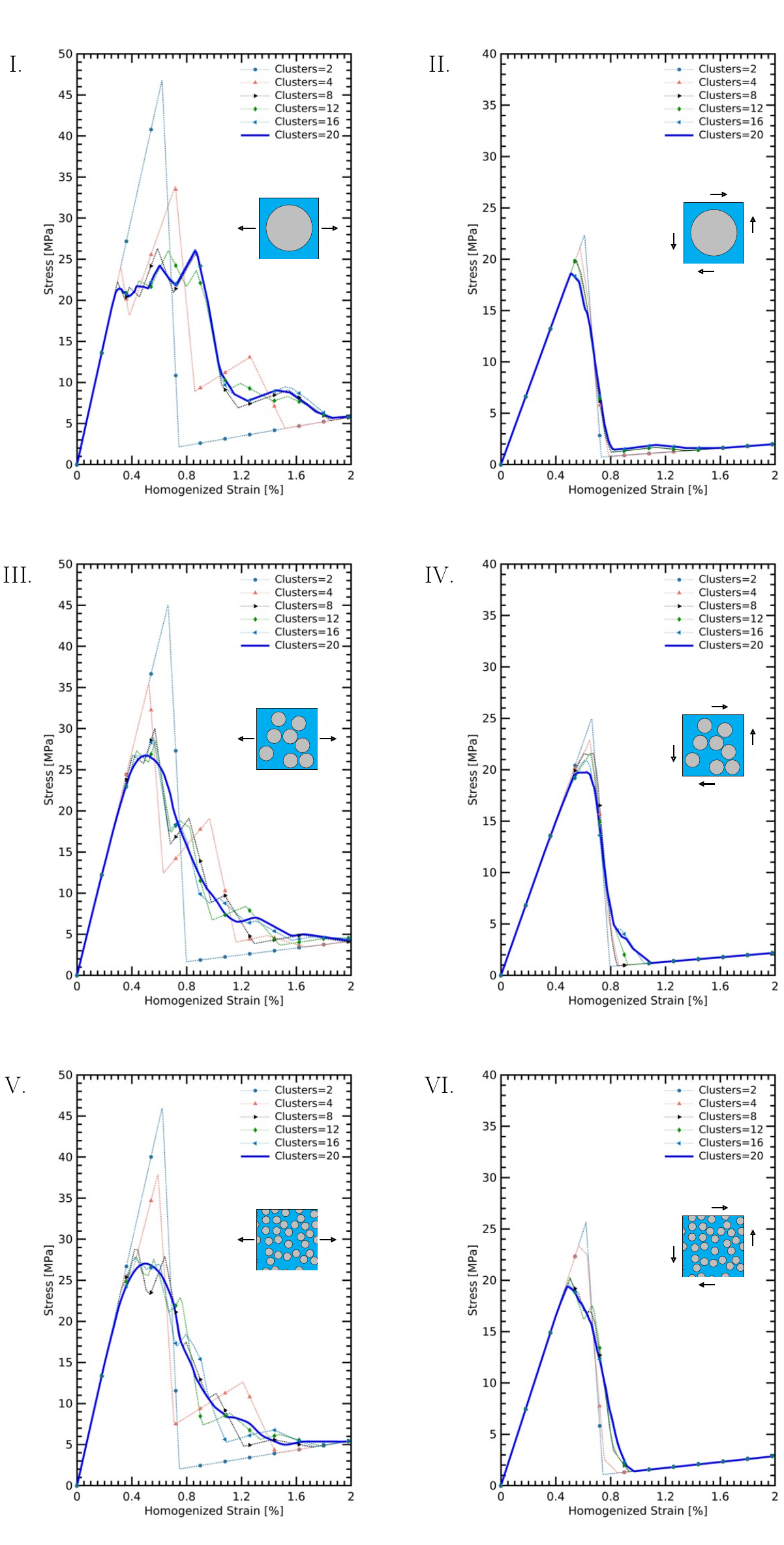}
\caption{Homogenized stress-strain response of single-fiber RVE (subfigures I and II), eight-fiber RVE (subfigures III and IV) and thirty-fiber RVE (subfigures V and VI) when subjected to tensile (subfigures I, III and V) and shear loading (subfigures II, IV and VI). The response is checked for different number of clusters ($\mathcal{M}=$ 2, 4, 8, 12, 16 and 20) obtaining by performing \textbf{eigen clustering scheme}}
\label{fig:plastic_cluster}
\end{figure}

Offline calculations are carried out to determine the partitioning map. Subsequently, the elastic and eigen influence tensors are calculated using Eq. (\ref{eq:49}) and (\ref{eq:55}). The clustering analysis is performed to divide the matrix into  $\mathcal{M}$ = 2, 4, 8, 12, 16 and 20 subdomains. Two different types i.e. elastic and eigen clustering schemes have been adopted for discretization. Periodic boundary conditions are applied to simulate the repeated microstructure and size independency. After the preprocessing stage, a finite-size homogenized domain is checked by applying tensile and shear loading using different number ($\mathcal{M}$ = 2, 4, 8, 12, 16 and 20) and types (elastic and eigen clustering) of partitions. Fig. \ref{fig:elastic_cluster} shows the variation of homogenized stress with applied strain for all three microscale domains using elastic clustering scheme. Fig. \ref{fig:plastic_cluster} presents the variation in homogenized stress as a function of applied strain for all three microscale domains and the data points were obtained by employing an eigen clustering scheme with cluster sizes ranging from $\mathcal{M}$ = 2 to $\mathcal{M}$ = 20.

Overall, the results highlight the sensitivity of mechanical behavior to microstructural arrangements and loading directions. These plots demonstrate variations in peak stress, softening, and failure strain for three different RVEs. For all the cases, the stress variation for 20 clusters, is smoother and more consistent compared to the fewer clusters, which exhibit higher fluctuations. Increasing the number of clusters (higher resolution) reduces stress variability and enhances the predictability of stress-strain responses. Smaller cluster counts lead to more noise in the curves, particularly in peak stress values. Higher inclusion density (Plots V and VI) leads to smoother stress-strain curves, with a tendency for lower peak stresses under vertical loading. No substantial improvement in the results was found when the number of partitions exceeded 20.     


Similar behavior has been observed for all RVEs when subjected to shear loading as shown in plots II, IV and VI of Fig. \ref{fig:elastic_cluster} and Fig. \ref{fig:plastic_cluster}. Furthermore, tensile loading shows higher peak stresses and failure strains compared to shear loading. Additionally, less variation of stress with respect to increase in clusters is examined in case of shear. Moreover, we observed that the single-fiber RVE fails to sufficiently represent the randomness of the microstructure, even when 20 partitions are used (as depicted in Fig. \ref{fig:elastic_cluster}.I and Fig. \ref{fig:plastic_cluster}.I). In contrast, the RVEs containing eight or thirty fibers consistently produced accurate predictions of the macroscopic response when partitioned using 20 number of clusters. Therfore, we conclude that for TFA-based homogenization of fiber-reinforced composites, RVEs containing multiple fibers are necessary to capture the essential characteristics of failure in the material behavior.\color{black} 

Besides, the comparision of results for eigen clustering (shown in Fig. \ref{fig:elastic_cluster}) with elastic (shown in Fig. \ref{fig:plastic_cluster}) evidences the smoother response of stress variation with same number of partitions. Section \ref{eigen_elastic} explains the further investigation carried out for comparison of eigen and elastic clustering schemes. \color{black}Moreover, it can be concluded that for fiber-reinforced composites with RVEs containing between 8 and 30 fibers, a partitioning into 20 clusters offers a good trade-off between accuracy and computational efficiency. Based on the stress-strain response, it has been decided to carry out further research using 30-fiber RVE implementing atleast 20 clusters. 

However, it is important to note that there is no universal or analytical method to determine the optimal number of clusters a priori for a random RVE. In practice, the optimal cluster count must be determined empirically by conducting preliminary simulations and observing the convergence behavior of the stress-strain response.\color{black}

\subsection{Numerical Implementation-2: Elastic v/s Eigen Clustering}
\begin{figure}[H]
\centering
\includegraphics[width=1\textwidth]{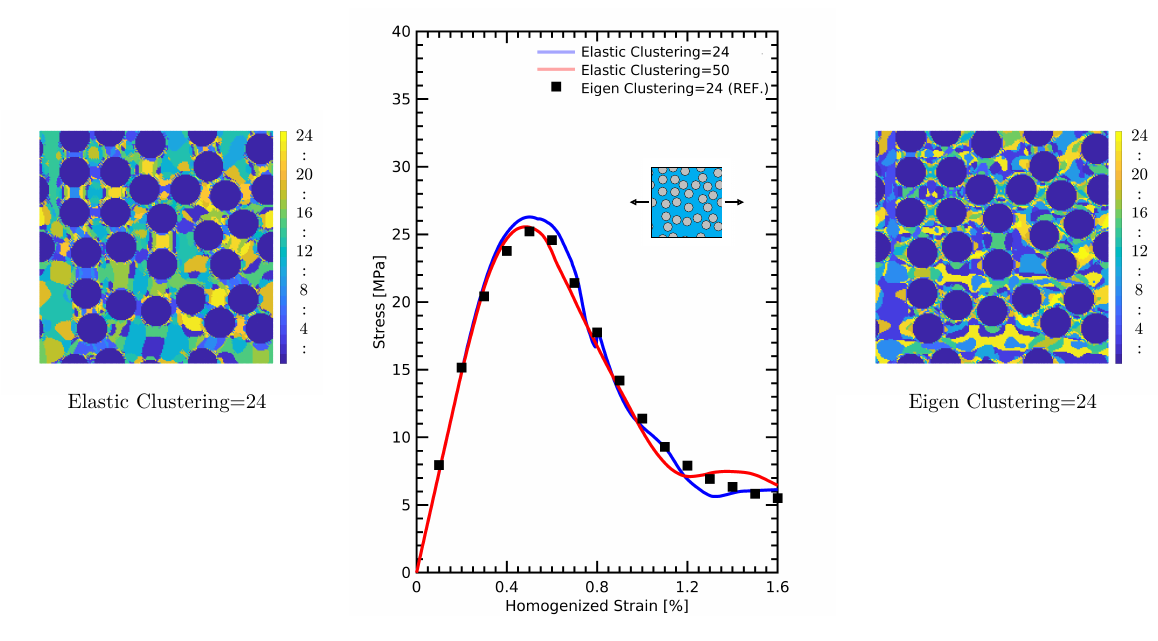}
\caption{Comparison of homogenized stress-strain response of RVEs with different number of partitions obtained from elastic and eigen clustering schemes. Each RVE contains thirty fibers and subjected to tensile loading. }
\label{fig:compare}
\end{figure}

{\label{eigen_elastic}}
In order to compare two partitioning schemes i.e. elastic and eigen clustering, a microscale analysis of RVE with 30 fibers is carried out under tensile loading. Fig. \ref{fig:compare} compares homogenized stress-stress behavior obtained using elastic and eigen clustering approaches. The stress variation with 24 number of eigen clusters is considered as a reference and the response of same number of elastic clusters is compared.

\vspace{0.5cm} 
Fig. \ref{fig:compare}  highlights the mismatch between the stress response of RVE with 24 number of elastic and eigen clusters. The number of elastic clusters are further increased and it is found that the results of 50 number of elastic clusters show improved agreement with eigen clustering in both elastic and softening regimes. Higher resolution allows for better stress and strain field representation, minimizing deviations. Hence, it is concluded that elastic low-resolution clustering (24 clusters) smooths stress fields significantly, underestimating local heterogeneities whereas the higher resolution (50 clusters) mitigates this issue and fully capture eigenstrain-induced effects during softening deformation. For high-fidelity modeling and minimum the computational cost, eigen clustering is preferable, particularly in failure-critical scenarios involving softening. 

\subsection{Numerical Implementation-3: Determination of Microscale Damage Morphology}
\begin{figure}[H]
\centering
\includegraphics[width=0.925\textwidth]{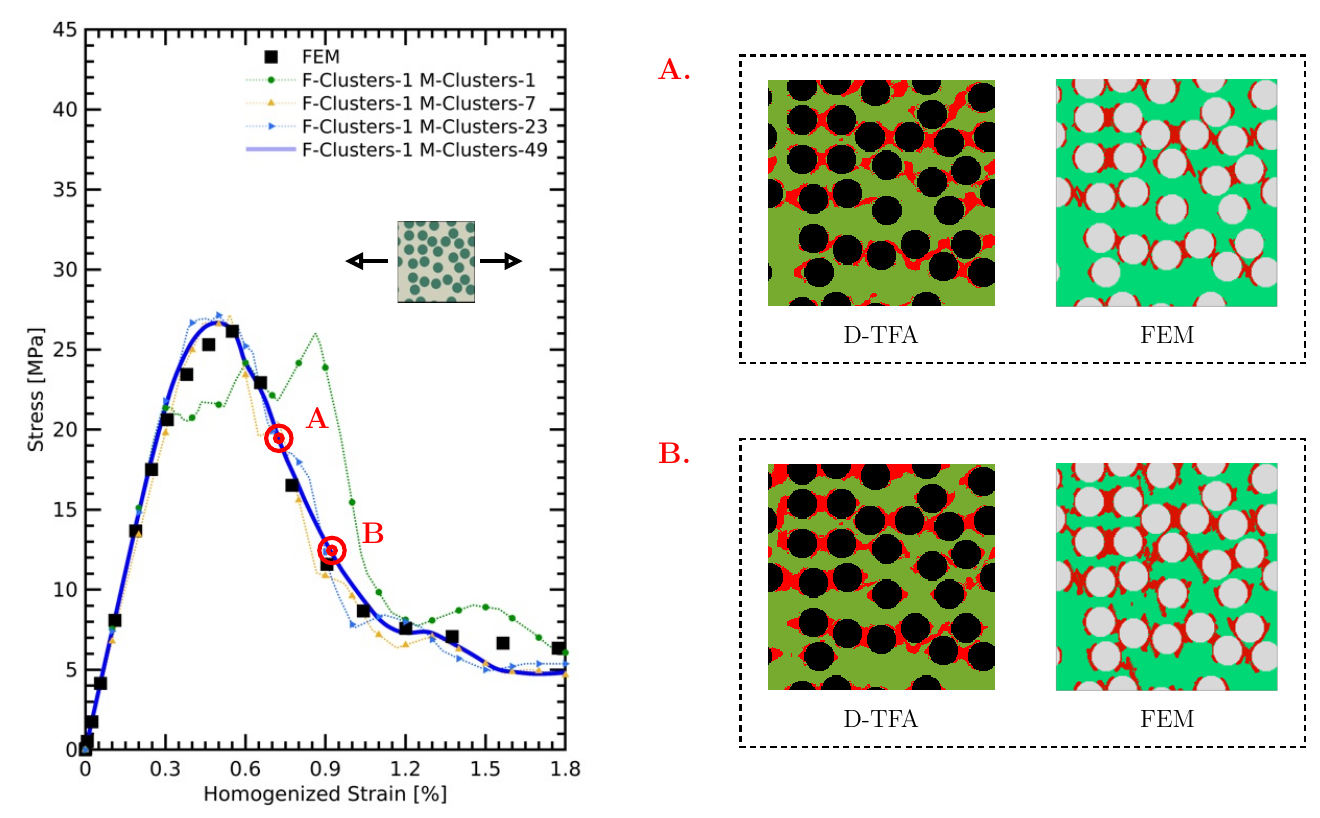}
\caption{Tensile stress-strain response of the homogenized medium for thirty-fiber RVE with different number of clusters ($\mathcal{M}=$ 2, 8, 24 and 50) and comparison with FEM when load is applied in $x$-direction. Damage morphologies obtained from D-TFA and FEM are compared at two points A. and B. located in the softening region. \color{black}In this context, "F-Clusters" designate partitions of the fiber phase, and "M-Clusters" refer to partitions of the matrix phase.}
\label{fig:damage_morph_tension_x}
\end{figure}

\begin{figure}[H]
\centering
\includegraphics[width=0.925\textwidth]{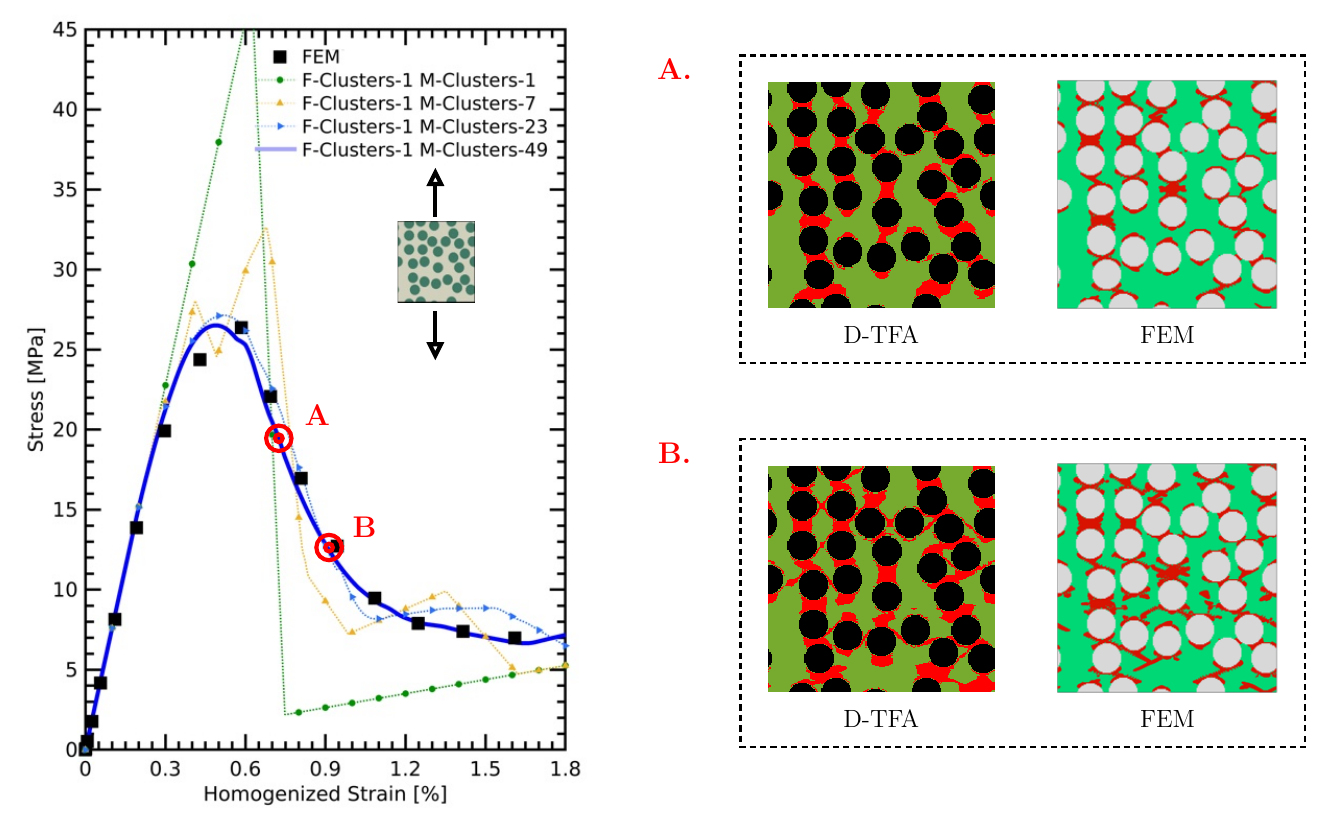}
\caption{Tensile stress-strain response of the homogenized medium for thirty-fiber RVE with different number of clusters ($\mathcal{M}=$ 2, 8, 24 and 50) and comparison with FEM when load is applied in $y$-direction. Damage morphologies obtained from D-TFA and FEM are compared at two points A. and B. located in the softening region. \color{black}In this context, "F-Clusters" designate partitions of the fiber phase, and "M-Clusters" refer to partitions of the matrix phase.}
\label{fig:damage_morph_tension_y}
\end{figure}

\begin{figure}[H]
\centering
\includegraphics[width=0.925\textwidth]{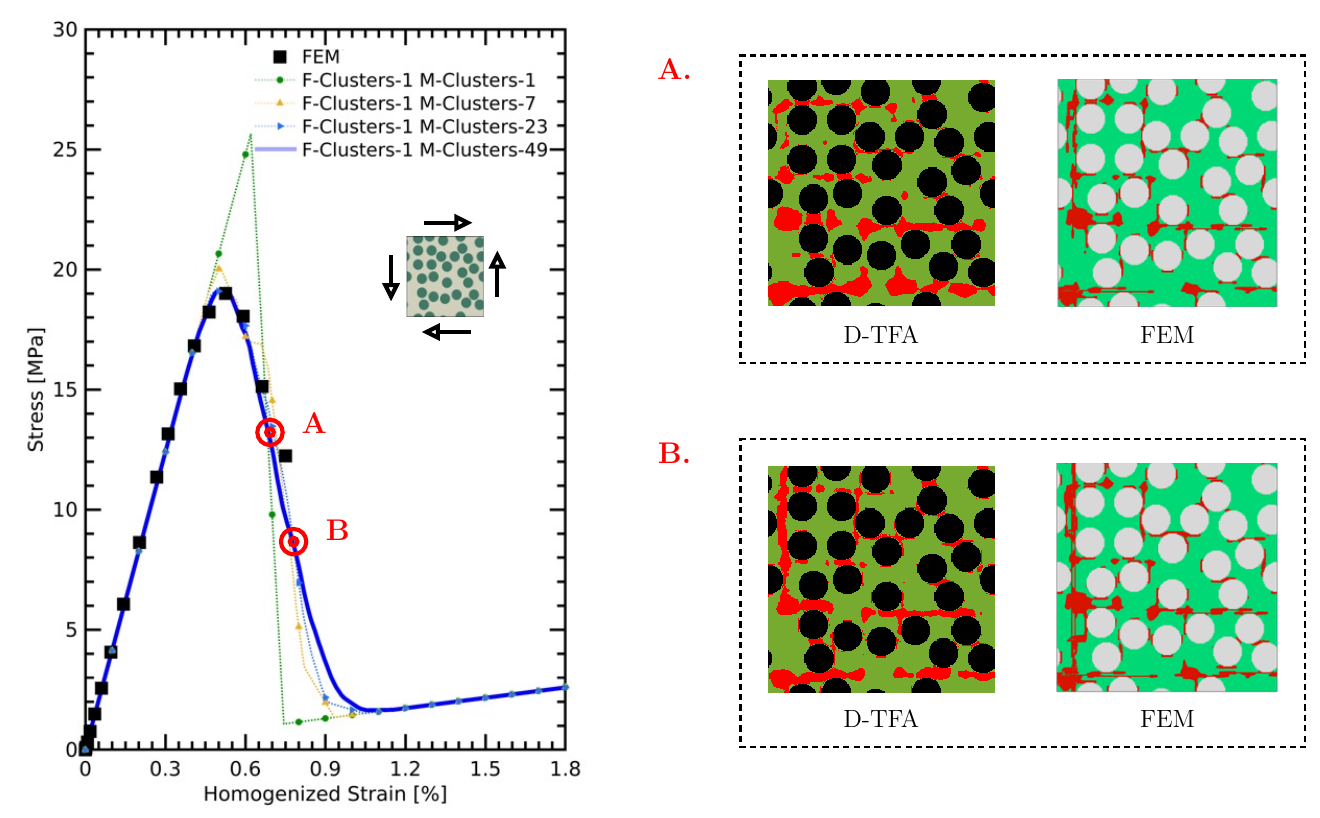}
\caption{Shear stress-strain response of the homogenized medium for thirty-fiber RVE with different number of clusters ($\mathcal{M}=$ 2, 8, 24 and 50) and comparison with FEM when shear load is applied in $xy$-plane. Damage morphologies obtained from D-TFA and FEM are compared at two points A. and B. located in the softening region. \color{black}In this context, "F-Clusters" designate partitions of the fiber phase, and "M-Clusters" refer to partitions of the matrix phase.}
\label{fig:damage_morph_shear_xy}
\end{figure}

Another study is carried out to investigate the damage morphologies obtained  through D-TFA and a 30-fiber RVE is explored when subjected to tension and shear. The homogenized stress-strain response and damage patterns are compared with FEM. Eigen clustering scheme is adopted for discretizing the microstructural domain into 2, 8, 24 and 50 eigen clusters. Fig \ref{fig:damage_morph_tension_x}, \ref{fig:damage_morph_tension_y} and \ref{fig:damage_morph_shear_xy} illustrate homogenized stress-strain behavior with a focus on damage morphologies calculated using D-TFA and clustering effects during damage-caused softening when subjected $x$-tensile, $y$-tensile and shear loadings respectively. It provides detailed insights into how clusters impact stress distribution, with snapshots (A and B) visualizing microstructural responses. The black squares represent FEM results, serving as a reference. Cluster configurations (e.g., 2, 8, 24, and 50 clusters) show how increasing resolution impacts stress-strain responses. Higher cluster numbers (50 clusters) result in smoother curves with better agreement to FEM, especially beyond the peak stress. The results are analyzed deeply at two points of interest (A and B) correspond to the interim state of full impairment of RVE and do not represent the fully damaged state of the material. Point A is located near the peak stress, indicating the onset of damage to the material \color{black}(macroscopic damage state of nearly 20\%)\color{black}. Point B corresponds to a softened state in the stress-strain curve, representing a significant loss of stiffness \color{black}(macroscopic damage state of approximately 70\%)\color{black}. These microstructural snapshots at points A and B compare the predictions of damage maps for D-TFA with FEM. The damage starts in the regions of the localized stress concentrations around inclusion edges. The red area represents the regions of matrix damage that match with FEM. At point B, the red zones spread wider, showing significant damage and interaction between inclusions. \color{black}These regions are basically the locations of stress concentrations from where the crack originates and propagates normal to the loading direction. Ultimately, various cracks merge into each other, which will be normal to the direction of applied loading. \color{black}Identical behaviour of stress-strain variations both in elastic and softening regimes validates D-TFA's predictions. \color{black}In composites with matrix materials exhibiting large softening, significant damage and deformation occur in the matrix surrounding a majority of fibers, especially in zones subjected to high local stress concentrations. Nevertheless, due to the heterogeneous distribution of stress at the microscale, not all matrix regions fracture uniformly. Some matrix zones, particularly those shielded by aligned fibers or located far from critical load paths, may still remain within the elastic regime as shown in Fig. \ref{fig:damage_morph_tension_x}, \ref{fig:damage_morph_tension_y} and \ref{fig:damage_morph_shear_xy}.\color{black} 

It is also observed that low clustering (2 Cluster) exhibits excessive fluctuations and premature softening, and high clustering (50 Clusters) produces smoother and more realistic stress-strain curves, closely matching FEM. This reflects better resolution in capturing inclusion-matrix interactions and damage-caused strain localization. \color{black} The stress–strain response displays an initial phase of sudden crack initiation, indicated by a sharp stress drop immediately after the peak load. This behavior corresponds to unstable fracture onset. Following this event, the crack propagates in a stable manner, as evidenced by a gradual reduction in stress with increasing strain. Such a response suggests the presence of post-initiation toughening mechanisms that provide resistance to rapid crack growth. \color{black}It is worth noting that the current methodology has proven effective for materials exhibiting ductile failure. Extending the proposed framework to analyze brittle fracture, particularly in glass-epoxy composites, is expected to be relatively straightforward. \color{black}

\color{black}

\vspace{0.5cm} 
\subsection{Numerical Implementation-4: 3D RVE Simulations for Tensile and Shear Characterization}
This numerical study explores the directional stress-strain response of a 3D microstructure under various loading conditions, comparing clustering resolutions (e.g., 2 clusters vs. 12 clusters). Essentially, the stress variation in the fiber direction under tensile loading and out of plane shear response are investigated using this microstructural study.

\begin{figure}
\centering
\includegraphics[width=1\textwidth]{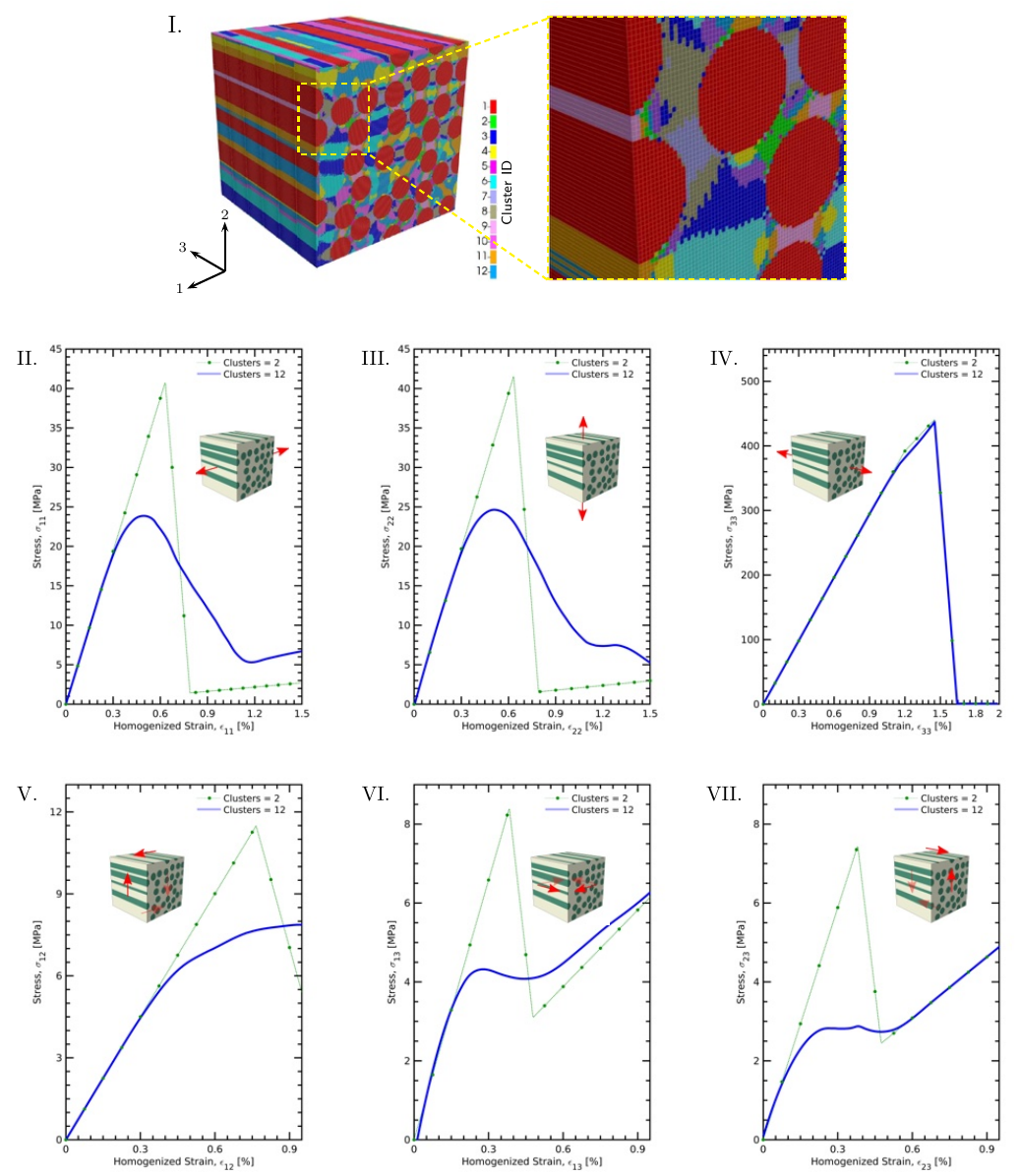}
\caption{The mechanical behavior of the homogenized medium for a 3-dimensional RVE with twelve partitions ($\mathcal{M} = 12$) is analyzed under various loading conditions: transverse tensile loading in the 1 and 2 directions (subfigures II and III), longitudinal tensile loading in the 3 direction (subfigure IV), in-plane shear (subfigure V), and out-of-plane shear (subfigures VI and VII). The response is also compared to that of a domain with two partitions.}
\label{fig:3d_rve}
\end{figure}

Fig. \ref{fig:3d_rve} depicts a 3D periodic microstructure with inclusions embedded in the matrix and each region is colored to represent different clusters. Eigen clustering scheme is implemented for microscopic domain discretization. The finite element domain is created using 968000 number of 8-noded hexahedral reduced integration elements. The RVE models a system consisting of multiple fibers embedded within a matrix. Material properties for each phase are detailed in Table \ref{table:RVE_properties}. The RVE is evaluated under six distinct loading conditions.

The analysis reveals that the stress-strain behavior exhibits strong directional dependency, demonstrating the importance of considering 3D anisotropic effects in material design. Furthermore, it is observed that higher clustering resolution (12 clusters) captures localized stress fields and anisotropic responses more effectively, whereas coarser clustering (2 clusters) is computationally cheaper but less reliable for complex, direction-dependent load cases. Elastic-to-damage transitions are smoother and more accurate with fine clustering, which is critical for predicting the directional failure mechanisms under intricate loading scenarios.  

\subsection{Numerical Implementation-5: Open Hole Tension Test Simulations}

\color{black}Open hole test simulations are performed to check the capability of the proposed methodology for predicting the response of the macroscopic domain when subjected to nonproportional loading conditions. These simulations involve multiaxial loading paths where the principal macrostrain axis of a local element rotates relative to its local coordinate system during load application. \color{black}In this example, the open-hole tension (OHT) test of a unidirectional (UD) composite plate, experimentally investigated by \cite{modniks2015analysis}, is revisited. This problem has been extensively studied in the literature, with researchers employing finite fracture mechanics (\citep{bleyer2018phase}) and phase-field approaches (\citep{bleyer2018phase}) to replicate the experimental results reported by \cite{modniks2015analysis}. These experimental and numerical investigations primarily focused on understanding the impact of fiber orientation on the effective strengths of the composite lamina plate. 

\vspace{0.5cm} 
\begin{figure}[H]
	\centering
	\includegraphics[width=1\textwidth]{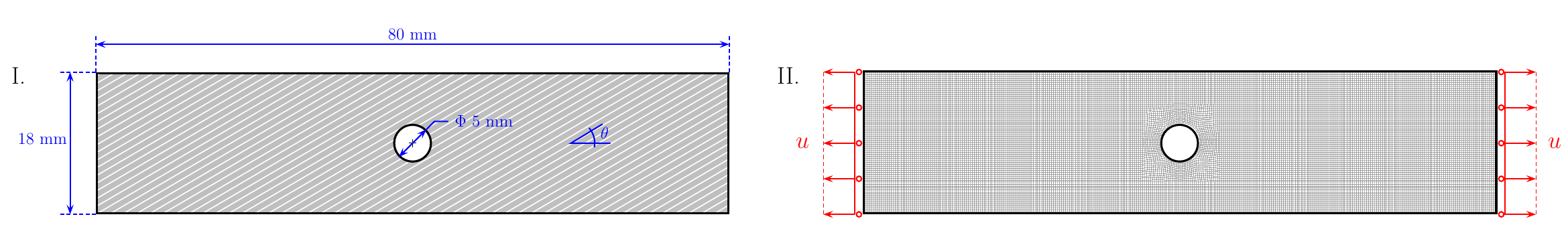}
	\caption{Illustration of the model used for simulating an open-hole specimen in tension:
		I. Geometric representation of a 80 × 18 plate featuring a central hole with a diameter of 2.5 mm; $\theta$ indicates the fiber layup direction.
		II. Finite element (FE) representation of the model, including boundary and loading conditions. All geometric dimensions are specified in millimeters}
	\label{fig:pwah_model}
\end{figure}

\vspace{0.5cm} 
The plate, featuring a circular hole with \( \text{diameter} = 5 \, \text{mm} \), has dimensions of \( \text{length} = 80 \, \text{mm} \), \( \text{width} = 18 \, \text{mm} \), and \( \text{thickness} = 0.1 \, \text{mm} \), as depicted in Fig. \ref{fig:pwah_model}. The fiber orientation relative to the horizontal direction is represented by \(\theta\). The material properties of the composite lamina are listed in Table~5.  fiber orientations of \(\theta = \{0^\circ, 30^\circ, 45^\circ, 60^\circ, 90^\circ\}\). Fig. \ref{fig:pwah_model} illustrates the detailed geometry, loading conditions, and boundary conditions for the lamina, where the fibers are oriented at an angle \(\theta\) relative to the global reference frame. The computational domain is discretized into 120,000 quadrilateral elements, with a mesh size of approximately \( 0.2 \, \text{mm} \) in the fracture process zone. The material properties used in the numerical simulations are provided in Table \ref{table:2}, sourced from Refs.~\cite{modniks2015analysis, felger2017mixed} which include the elastic properties and strength parameters for different phases.   
\begin{table}[h]
	\captionsetup{width=0.75\textwidth}
	\caption{Material properties for the phase materials - flax and epoxy used for open hole test simulations. Source: \cite{modniks2015analysis, felger2017mixed}}
	
	\centering
	\begin{tabular}{lcc}
		\toprule
		\textbf{Material Property} & \textbf{Fiber }{- Flax} & \textbf{Matrix }{- Epoxy} \\
		\midrule
		Volume Fraction & 0.41 & 0.59\\
		Elastic Modulus [MPa]& 43,050 & 2,670 \\
		Poisson's Ratio & 0.3 & 0.3 \\
		Damage Initiation Strain [\%] & 1.40 & 2.27 \\
		Damage Failure Strain [\%] & 2.80 & 4.55 \\
		\bottomrule
	\end{tabular} \label{table:2}
\end{table}

\color{black}Initially, an RVE containing eight fibers is utilized during the preprocessing stage to compute the elastic and eigen influence tensors. Furthermore RVE simulations are performed for obtaining new partitioning map was generated using the $k$-means clustering algorithm, tailored to the specific mechanical response of this updated RVE configuration. Given that the material properties of the constituent phases in this case differ from those used in earlier simulations, an eigenstrain based new partitioning map is generated using the $k$-means clustering algorithm.

\begin{figure}[H]
	\centering
	\includegraphics[width=0.965\textwidth]{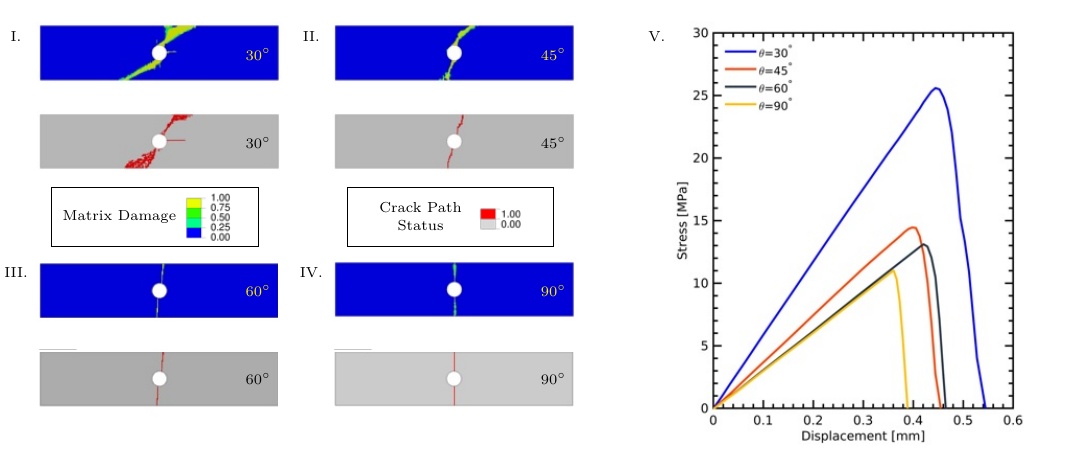}
	\caption{Structural behavior of the open-hole specimen under tensile loading, represented through matrix damage variables and fracture path contours, is shown for fiber layup directions of I. 30$^\circ$, II. 45$^\circ$, III. 60$^\circ$, and IV. 90$^\circ$. Subfigure V compares the stress-displacement behavior across all four cases.}
	\label{fig:pwah-2}
\end{figure}

\begin{figure}[H]
	\centering
	\includegraphics[width=0.965\textwidth]{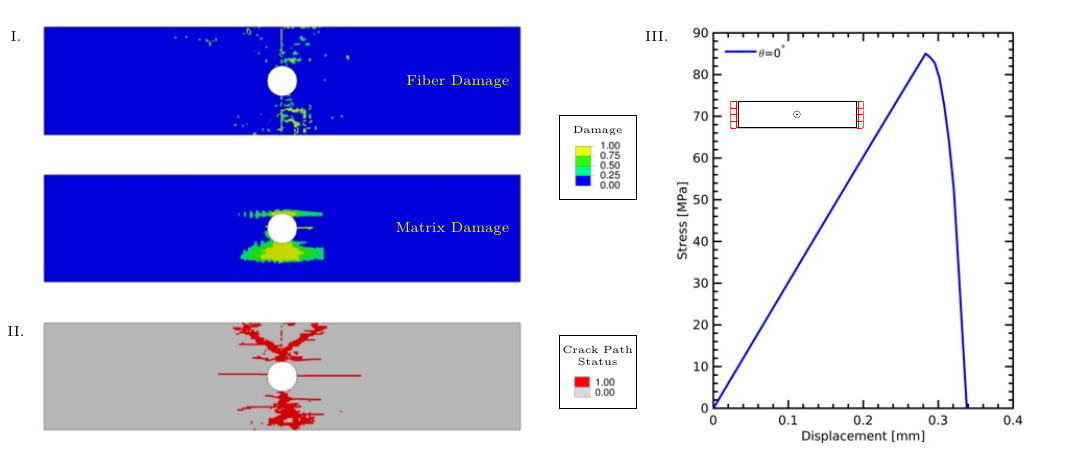}
	\caption{The mechanical response of the 0$^\circ$ open-hole specimen under tensile loading is presented in terms of I. fiber and  matrix damage variables contours. Subfigure II depicts the resulted fracture path contour and subfigure III illustrates the stress-displacement plot.}
	\label{fig:pwah-3}
\end{figure}

Once the influence tensors are computed, the macroscale simulations are performed following the methodology detailed in Section \ref{sec:method}. \color{black}The predicted crack patterns for different material orientations are depicted in Fig. \ref{fig:pwah-2} I, II, III, IV and the associated stress-displacement curves are given in Fig. \ref{fig:pwah-2} V. Damage morphologies are estimated by plotting the contours of variables for 1). matrix damage, $\bar{{\mathbf{D}}}_{\sigma}$, 2). fiber damage, $\bar{{\mathbf{D}}}_{\epsilon}$ and 3). overall damage, $\bar{{\mathbf{D}}}$ which corresponds to crack in the material, as shown in Fig. \ref{fig:pwah-2} and \ref{fig:pwah-3}. Moreover, it is seen that the predicted crack paths are parallel to the fiber orientation in all the cases, which is in line with the experimental observation as reported by \cite{modniks2015analysis}. \color{black}These cracks start at the location of high stress concentration near the hole and propagate along the fiber directions till they reach to the edges of specimen. As soon the crack propagates to the edge, material loses its complete load-carrying capacity. For these specimens, it is observed that the complete failure happens due to matrix damage. \color{black}

 Fig. \ref{fig:pwah-3} I and II presents the multiple crack evolutions of the 0 ply at different load levels as marked in the stress-displacement curve of Fig. \ref{fig:pwah-3} III. It is seen that, the matrix splitting crack initiates at the corners of hole at the load level of about 20 MPa and propagate gradually towards the specimen edge. We further increase the load to obtain the ultimate strength and final crack pattern of the plate. As can be appreciated from the figure, the fiber cracks initiate around the hole at nearly 95\% of the ultimate load and propagates rather quickly towards the upper and lower edges of the specimen. The rapid fiber rupture leads to a complete failure with two-part separation of the plate, which is also demonstrated by the sudden load drop of the force-displacement curve.   \color{black}Furthermore, the failure of fibers will lead to instability in the distribution of forces around the hole and result unsymmetric distribution of the stress field in the overall domain. This disturbance causes a slight disproportionate distribution of matrix and fiber damage which results the zig-zag pattern of the crack/damage map (shown red in Fig. \ref{fig:pwah-3} II). \color{black}
 
 Additionally, in strain-softening systems, damage localization acts as a bifurcation phenomenon, minor perturbations such as round-off errors, geometric imperfections, or mesh asymmetries can trigger asymmetric damage evolution, even under macroscopically symmetric loading. In numerical simulations, additional sources of asymmetry can arise from discretization artifacts. Even when geometric and loading conditions are symmetric, small imperfections in the mesh or numerical noise can serve as triggers for bifurcation. In damage simulations, these imperfections manifest as non-uniform strain localization, leading to macroscale asymmetry in the damage field. \color{black}

\begin{figure}[H]
	\centering
	\includegraphics[width=1\textwidth]{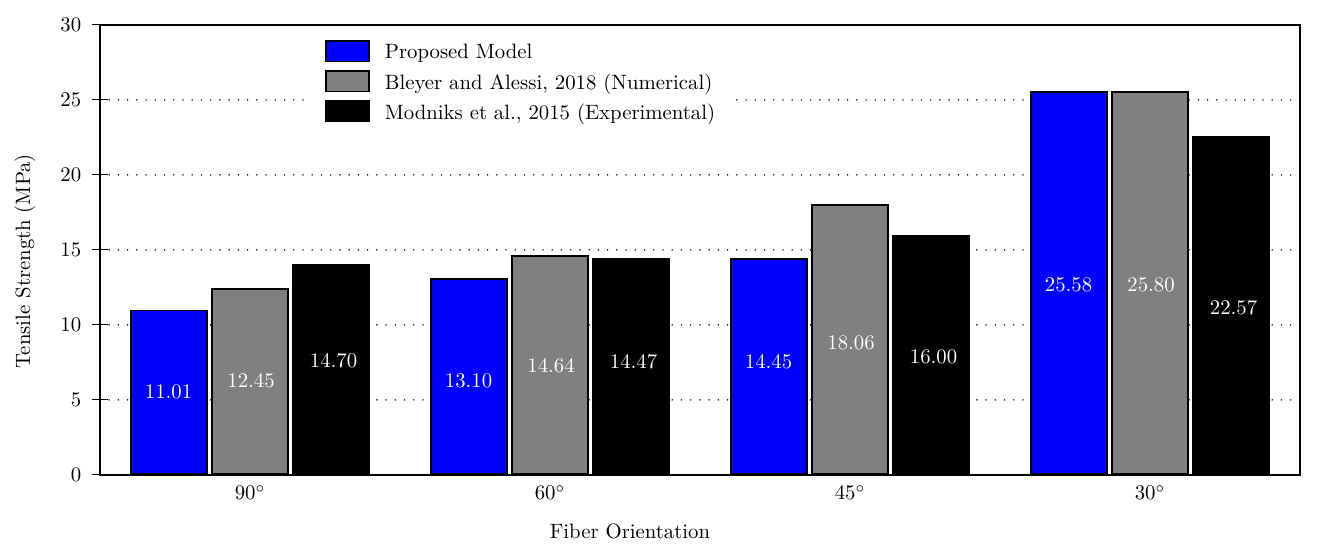}
	\caption{Comparison of calculated tensile strengths of 90$^\circ$, 60$^\circ$, 45$^\circ$, and 30$^\circ$ open hole test specimens with numerical predicted results by \cite{bleyer2018phase} and experimentally determined values by \cite{modniks2015analysis}} 
	\label{fig:pwah-4}
\end{figure}

Furthermore, the effective strength of the plate was calculated using D-TFA approach and compared with experimental data and numerical predictions obtained by other technique, as reported in references \cite{modniks2015analysis} and \cite{bleyer2018phase}. To validate the accuracy of the proposed method, Fig. \ref{fig:pwah-4} presents a comparison of the numerically predicted and experimentally observed effective strengths of the composite lamina. A significant increase in effective strength is observed when the fiber orientation exceeds \(\theta > 30^\circ\), aligning with the experimentally observed trend. For orientations of \(\theta \leq 30^\circ\), the fiber contributes minimally to the load-carrying capacity. The predicted failure strengths at various fiber orientations show reasonable agreements with other numerical formulation proposed by \citep{bleyer2018phase}, as shown in Fig. \ref{fig:pwah-4}. Note that this nearly dynamic, collapse fracture process is successfully captured by the proposed D-TFA. 

\color{black}This comparison serves to highlight that the proposed D-TFA-based multiscale framework delivers a level of predictive accuracy that is on par with other established multiscale modeling techniques found in the literature. \color{black} It is well-documented in the literature that conventional TFA-based approaches often suffer from reduced predictive accuracy, particularly under complex loading or damage evolution scenarios. In this context, the present comparison underscores that the proposed damage-enhanced TFA (D-TFA) multiscale framework achieves a predictive capability comparable to other established multiscale modeling techniques reported in the literature. For example, \cite{bleyer2018phase} developed a phase-field-based multiscale damage framework that demonstrates high fidelity but is computationally intensive, both in terms of time and implementation complexity. In contrast, the proposed D-TFA methodology offers significant computational efficiency. Once the preprocessing stage is completed, macroscopic simulations can be performed with minimal computational cost and reduced runtime, enabling efficient multiscale analysis without compromising accuracy. 
\color{black}

\section{Conclusions}\label{sec:7}

This work introduces D-TFA, a homogenization method for analyzing the mechanical behavior of composite materials with damage. D-TFA uses a novel continuum damage mechanics (CDM) framework to evaluate the microscopic eigenstrain field, which accounts for intra-phase damage. Damage evolution is modeled through rate equations for internal variables, derived from a dissipation potential. The method efficiently calculates preprocessing data, including elastic and eigen influence tensors ($\bar{\mathbf{E}}^{(i)}$ and $\bar{\mathbf{S}}^{(ij)}$), and homogenized damage-informed constitutive tensors ($\bar{\mathbf{L}}_d$ and $\bar{\mathbf{M}}_d^{(i)}$) by assuming a piecewise constant eigenstrain field in microscale partitions. Elastic and eigen clustering schemes enhance computational efficiency within a multiscale framework. Critically, clustering mitigates post-damage stiffness issues common in traditional TFA methods when applied to softening materials. D-TFA is a two-stage multiscale homogenization technique validated across various composite simulations. Initially, 2D RVE studies demonstrated its accuracy in predicting microscale response and eliminating pseudo-stiffness, benefiting from simplified preprocessing. It accurately modeled complex 2D damage trajectories and determined 3D RVE properties under multiple loads. Finally, D-TFA successfully predicted crack paths in open-hole laminate tensile tests, showing superior accuracy compared to other methods and strong agreement with experimental data.

\color{black}Traditional TFA-based methods struggle with determining the appropriate number of reduced-order variables, which are essential for accurate and efficient modeling. In the present work, we proposed a $k$-means-driven clustering method for deciding the overall partitioning map for the microscale domain. We concluded that out of two partitioning schemes, i.e. elastic v/s eigen clustering, the eigen clustering scheme proves more efficient. The effect of changing the partitions was also explored, which confirms that a smaller number of clusters is required in the case of the eigen clustering scheme than the elastic clustering scheme. Although it is also shown that twenty(20) partitions are adequate for capturing the accurate stress-strain response of RVE, it also requires that the RVE domain should contain a sufficient number of inclusions/fibers. It is observed that a single fiber RVE is incapable of capturing the behaviour of the RVE, even with a large number of partitions. A well-documented limitation of lower-order TFA models is the unrealistic residual stiffness observed in the RVE following damage under certain loading conditions, especially transverse loading. In the proposed multiscale framework, this challenge is alleviated by introducing a novel method for calculating macroscopic damage variables for uniform stress and uniform strain fields. RVE under transverse loading constitutes the uniform stress condition, whereas longitudinal loading along fibers exhibits the condition of uniform strain. An auxiliary stress field driven formulation is devised to apply a uniform stress field condition, and by comparing the statically admissible microscale fields, macroscopic damage is calculated. This results in almost zero stiffness of the RVE at the fully damaged state. Furthermore, the damage tensor for generalized multiaxial loading conditions is also presented, which gives an accurate representation of the stress field at macroscopic level.   
          
In summary, the D-TFA framework not only preserves the core advantages of traditional TFA-based multiscale methods but also addresses two significant limitations through innovations in clustering strategy and damage modeling. These developments significantly improve the accuracy and applicability of multiscale simulations for composite materials under diverse loading conditions. The proposed framework mainly handles the problems associated with fiber-reinforced composites when subjected to rate-independent loading conditions. It is particularly well-suited for simulating heterogeneous material domains composed of isotropic elastic phases that undergo isotropic damage evolution. The framework accurately captures the macroscale response of such composites by leveraging microscale information, making it a reliable tool for a wide range of mechanical applications. \color{black} While the current D-TFA formulation uses an isotropic damage representation, future work will address its limitations.  Specifically, the model does not currently account for the directional nature of micro-cracks.  Anisotropic damage models, as discussed in several publications \citep{levasseur2011two, fritzen2011nonuniform, morin2017micromechanics, pei2022anisotropic, ren2023micro, liu2023anisotropic}, offer a more realistic representation of damage and will be incorporated into D-TFA in the future.  Additionally, rate-dependent effects, crucial for capturing the viscoelastic behavior of fiber-reinforced composites under varying strain rates (important for dynamic applications), will be integrated.  

\color{black}D-TFA is suitable for both monotonic and cyclic loading. In this work, we demonstrated the proposed model's strong prediction capability using an RVE under monotonic loading. Many composites, however, experience permanent deformations and residual strains under cyclic loads, often leading to material failure. Currently, D-TFA incorporates damage-induced eigenstrains as internal variables to simulate microstructural degradation and effectively represent failure progression from material damage. However, it doesn't yet account for other sources of eigenstrains, such as those from plastic deformation, thermal expansion, or residual stresses. Future enhancements will expand the framework to include plastic and thermal eigenstrains, allowing its application to a broader range of materials and both monotonic and cyclic loading conditions.\color{black}

Finally, although D-TFA effectively models intra-phase damage, future research will focus on incorporating interface damage modes, which are currently under investigation by the authors. Despite these limitations, D-TFA remains a valuable multiscale approach for efficiently and accurately predicting the overall behavior of composite materials.   

\section*{Acknowledgement}
This work was supported by a research grant (ARDB/01/1051983/M/I, Project No. 1983) of the Aeronautics Research \& Development Board, India. The content is solely the responsibility of the authors and does not necessarily represent the official views of the funding organisation. \newpage

\bibliography{bibfile1}
\end{document}